\documentclass[prd,amsmath,amssymb,nofootinbib]{revtex4}
\usepackage{epsfig}
\usepackage{textcomp}
\usepackage{epstopdf}

\def\beq{\begin{equation}}
\def\eeq{\end{equation}}
\def\bea{\begin{eqnarray}} 
\def\eea{\end{eqnarray}}
\def\nn{\nonumber}
\def\gev{{\rm GeV}}
\def\tev{{\rm TeV}}
\def\mev{{\rm MeV}}
\def\kev{{\rm keV}}

\def\fb{{\rm fb}}
\def\pb{{\rm pb}}
\def\eps{\varepsilon}

\def\br{{\tt BR}}

\newcommand{\lsim}{
\mathrel{\hbox{\rlap{\hbox{\lower4pt\hbox{$\sim$}}}\hbox{$<$}}}}
\newcommand{\gsim}{
\mathrel{\hbox{\rlap{\hbox{\lower4pt\hbox{$\sim$}}}\hbox{$>$}}}}
\newcommand{\mat}[2][ccccc]{\left( \begin{array}{#1} #2\\ \end{array}\right)}

\begin{document}

\title{Dark Two Higgs Doublet Model}
\author{Hye-Sung Lee$^{1,2}$}
\author{Marc Sher$^1$}
\affiliation{
$^1$Department of Physics, College of William and Mary, Williamsburg, VA 23187, USA\\
$^2$Theory Center, Jefferson Lab, Newport News, VA 23606, USA}
\date{March 2013}
\begin{abstract}
We perform a detailed study of a specific Two Higgs Doublet Model (2HDM) with a $U(1)$ gauge symmetry, instead of a typical $Z_2$ discrete symmetry, containing a very light gauge boson $Z'$ (GeV scale or below).
The Standard Model (SM) fermions do not carry $U(1)$ charges, but induced couplings to the $Z'$ (called the dark $Z$) are generated through mixing with the SM neutral gauge bosons.
Such a light $Z'$ could explain some astrophysical anomalies as well as the muon $g-2$ deviation, and has been the subject of great experimental interest.
We consider the scenario in which the $125 ~\gev$ SM-like Higgs ($H$) is the heavier scalar state, and focus on the lighter neutral state ($h$) as well as charged Higgs.
We analyze the constraints on the model from various experiments and predict novel channels to search for these Higgs scalars at the LHC.
In particular, experiments looking for lepton-jets are among potentially important searches.
\end{abstract}
\maketitle

\section{Introduction}
At long last, the experimental exploration of the Higgs sector of the electroweak theory has begun.   The discovery of a scalar boson with a mass of approximately $125$ GeV at the Large Hadron Collider (LHC) is a spectacular triumph of the Standard Model (SM) \cite{:2012gk,:2012gu}.
The hints of a larger than expected branching fraction for $H \to \gamma\gamma$, although they have recently weakened, could suggest possible extensions of the scalar sector \cite{Ferreira:2011aa} or of other sectors.
(For some vector-like fermion extension examples, see Refs.~\cite{Dawson:2012di,Carena:2012xa,Bonne:2012im,An:2012vp,Joglekar:2012vc,ArkaniHamed:2012kq,Almeida:2012bq,Kearney:2012zi,Ajaib:2012eb,Davoudiasl:2012ig,Dawson:2012mk,Davoudiasl:2012tu}.)
Certainly, detailed analyses of these extensions are worthwhile and might provide guidance to experimenters searching for physics beyond the SM.  

The simplest and most studied such extension is the Two Higgs Doublet Model (2HDM), recently reviewed in Ref.~\cite{Branco:2011iw}.     A potential problem for such models are tree-level flavor changing neutral currents (FCNC) which will occur whenever fermions of a given charge couple to more than one Higgs multiplet \cite{Glashow:1976nt,Paschos:1976ay}.   While it is possible to ameliorate the constraints by assumptions about the flavor-changing couplings \cite{Cheng:1987rs, Pich:2009sp}, the usual solution to the problem is eliminating the tree-level FCNC altogether by using a discrete $Z_2$ symmetry.      Depending on the transformation properties of the right-handed fermions under the $Z_2$, several different models can be obtained \cite{Branco:2011iw}.   The most familiar two models are the Type I model, in which all of the fermions couple to a single Higgs multiplet, and the Type II model, in which the down-type quarks and charged leptons couple to one Higgs multiplet and the up-type quarks couple to the other.

A $Z_2$ symmetry can be promoted to a gauged $U(1)$ symmetry as discussed in Refs.~\cite{Davoudiasl:2012ag,Ko:2012hd}.
In Ref.~\cite{Davoudiasl:2012ag}, the SM fermions do not have charges of the new $U(1)$ gauge symmetry and a Type I 2HDM model results.
(See Ref.~\cite{Lee:2012xn} for another example in the fourth generation context.)
In Ref.~\cite{Ko:2012hd}, the SM fermions have charges under the $U(1)$ gauge symmetry, resulting in various types of 2HDMs.
As emphasized in Ref.~\cite{Ko:2012hd}, one of the interesting features in this kind of model is the absence of the pseudoscalar $A$.
It is a common effect \cite{Barger:2006dh} of a spontaneously broken $U(1)$ gauge symmetry that the pseudoscalar is eaten to become the longitudinal component of the new gauge boson $Z'$.
Of course, one could restore the pseudoscalar by adding a complex Higgs singlet to the model.   

In our paper, we will closely follow the scenario of  Ref.~\cite{Davoudiasl:2012ag}.
In this model, all SM fermions are neutral under the new Abelian gauge group $U(1)'$, yet the $Z'$ coupling to the SM fermions can be mediated by mixing with the SM neutral gauge bosons.
The $Z'$ can communicate with the SM fermions only through kinetic and/or mass mixing with the SM neutral gauge bosons.
The fact that the $Z'$ does not directly interact with fermions eliminates the numerous constraints on $Z'$ masses, and allows the $Z'$ to be extremely light, with mass allowed below ${\cal O}(1) ~\gev$.
(For a review on a complementary heavy $Z'$, see Ref.~\cite{Langacker:2008yv}.)

This model in Ref.~\cite{Davoudiasl:2012ag} was not originally proposed in the context of 2HDMs.
It was first proposed as a generalization of the so-called dark photon model whose coupling is of the same form as the photon coupling.
While the two models both adopt  kinetic mixing of the $U(1)_Y$ and the $U(1)'$, the difference comes from the fact that the $Z'$ gets its mass only from a Higgs singlet in the dark photon model while it gets the mass from a Higgs doublet (and also from a Higgs singlet if a Higgs singlet exists) in the dark $Z$ model.
As a result, the dark $Z$ can couple to the weak neutral current as well.
When the $Z'$ is very light, it can explain \cite{Boehm:2003bt,Boehm:2003ha,ArkaniHamed:2008qn} the astrophysical anomalies such as the $511 ~\kev$ gamma-ray from the Galactic center observed by the INTEGRAL satellite \cite{Jean:2003ci} or the positron excess observed by ATIC \cite{Chang:2008aa} and PAMELA experiments \cite{Adriani:2008zr}, depending on the property of the dark matter.
It can also explain the $3.6 \sigma$ deviation of the measured muon $g-2$ from the SM prediction \cite{Bennett:2006fi,Beringer:1900zz}.
(For some discussions about non-Abelian dark gauge sectors, see Refs.~\cite{Baumgart:2009tn,Chen:2009ab,Zhang:2009dd,Carone:2013uh}. See also Ref.~\cite{Alanne:2013dra} for a discussion on $W$-$W'$ mixing.)

It is remarkable that a very light gauge boson, introduced to provide a dark matter explanation to some astrophysical anomalies, can also provide an explanation for the absence of tree-level FCNC in a 2HDM.   It leads to a very different phenomenology from the ordinary 2HDMs.   Some of the implications for the dark $Z$ boson as well as the $125 ~\gev$ SM-like Higgs state, $H$,  of this model have been discussed in Refs.~\cite{Davoudiasl:2012ag,Davoudiasl:2012qa,BNL}.
In this paper, we will mainly discuss the phenomenology of the other Higgs bosons of the model, and refer to this model as the ``Dark 2HDM'' to emphasize that our study of this model focuses on the comparison to the ordinary 2HDMs.
We particularly focus on the scenario that the other Higgs, $h$,  is lighter than the SM-like Higgs, $H$, of $125 ~\gev$, which was a subject of the study in Ref.~\cite{Ferreira:2012my} for the ordinary 2HDMs.    We will see that various constraints force one to add a singlet to the model.   The decay of the $h$ to $Z'Z'$ will provide a dramatic signature that could be on the verge of discovery at the LHC, and the decay of the charged Higgs to $h W \to Z'Z'W$ would also lead to striking signatures.

In Sec.~\ref{sec:model}, we describe the model and the particle masses and couplings.
In Sec.~\ref{sec:constraints}, we consider the constraints from LEP and the LHC on the neutral Higgs bosons, and in Sec.~\ref{sec:lightHiggs} study the remarkable signature of the light Higgs and the implications for LHC experiments looking for multi-lepton jets.
In Sec.~\ref{sec:chargedHiggs}, we discuss the properties of the charged Higgs boson and
in Sec.~\ref{sec:conclusion}, we present our conclusions.
A discussion of the properties of the $Z'$ boson is in the Appendix.

\section{The Model}
\label{sec:model}
In this section, we describe the Dark 2HDM proposed by Davoudiasl, Lee, and Marciano \cite{Davoudiasl:2012ag}.
The scalar sector is composed of two doublets and a singlet under the $SU(2)_L$.
As we will see, the model with just two doublets and no singlet is not compatible with various constraints.
Including a singlet can avoid this issue.

The gauge group is $SU(3)_C\times SU(2)_L \times U(1)_Y \times U(1)'$, and the SM fermions are assumed to be neutral under the extra $U(1)'$.      The gauge part of the kinetic Lagrangian includes kinetic mixing between the $U(1)_Y$ and $U(1)'$ \cite{Holdom:1985ag} and is given by
\beq
{\cal L}_{\rm gauge} = -\frac{1}{4} \hat B_{\mu\nu} \hat B^{\mu\nu} + \frac{1}{2}\frac{\eps}{\cos\theta_W} \hat B_{\mu\nu} \hat Z^{'\mu\nu} -\frac{1}{4} \hat Z'_{\mu\nu} \hat Z^{'\mu\nu} \label{eq:kineticMixing}
\eeq
where $\hat B_{\mu\nu}\equiv \partial_\mu \hat B_\nu - \partial_\nu \hat B_\mu$ and $\hat Z'_{\mu\nu} \equiv \partial_\mu \hat Z'_{\nu} - \partial_\nu \hat Z'_{\mu}$.
The hat notation indicates the states before diagonalization.
As discussed in various places, for examples see Refs.~\cite{Gopalakrishna:2008dv,Davoudiasl:2012ag}, one can redefine the fields to remove the kinetic mixing term parametrized by $\eps$, which is experimentally constrained to be small.

When there is one Higgs doublet and the $Z'$ gets the mass from a Higgs singlet, this leads to an induced coupling of the $Z'$ to the electromagnetic current
\beq
{\cal L}_\text{dark $\gamma$} = -\eps e J^\mu_{em} Z'_\mu \qquad \left(J^\mu_{em} \equiv Q_f \bar f \gamma^\mu f\right)
\eeq
and the induced coupling of the $Z'$ to the weak neutral current is negligibly small because of the suppression by high orders of $\eps$.
This gauge boson is widely called the ``dark photon'', since fermions couple to the $Z'$ with a very small effective electric charge ($\eps Q_f$) for a given electric charge ($Q_f$) of the fermion $f$.

The model in Ref.~\cite{Davoudiasl:2012ag} added an additional Higgs doublet charged by the $U(1)'$  which introduces mixing in the $Z$-$Z'$ mass matrix.
The $Z$-$Z'$ mass mixing, parametrized by $\eps_Z$, is also taken to be small.
The value of $\eps_Z$ will be determined by parameters in the scalar potential.
Because of this mixing, the $Z'$ couples to both electromagnetic current and weak neutral current as
\beq
{\cal L}_\text{dark $Z$} = -\left(\eps e J^\mu_{em} + \eps_Z g_Z J^\mu_{NC} \right) Z'_\mu \qquad \left(J^\mu_{NC} \equiv [\frac{1}{2} T_{3f} - Q_f \sin^2\theta_W] \bar f \gamma^\mu f - [\frac{1}{2} T_{3f}] \bar f \gamma^\mu \gamma_5 f\right)
\eeq
with $g_Z = g / \cos\theta_W = g' / \sin\theta_W \simeq 0.74$, $T_{3f} = \pm 1/2$ and the weak mixing angle $\sin^2 \theta_W \simeq 0.23$ .
The $Z'$ with more general coupling (compared to the dark photon) in this model was named  the ``dark $Z$'' to emphasize that it couples to the weak neutral current.
In particular, when the $\eps = 0$ limit is taken, the $Z'$ couples only to the weak neutral current as the SM $Z$ boson does with a suppressed coupling ($\eps_Z g_Z$).
The definition of $\eps_Z$ and constraints on  $m_{Z'}$, $\eps$, $\eps_Z$ can be found in the Appendix.

We use a similar but slightly different notation from Ref.~\cite{Davoudiasl:2012ag}.\footnote{In Ref.~\cite{Davoudiasl:2012ag}, the field $\Phi_1$ was chosen to couple to the SM fermions rather than $\Phi_2$, with definitions of $\tan\beta \equiv v_2 / v_1$, $\tan\beta_d \equiv v_2 / v_S$.
This leads to different typical values of $\tan\beta$ and $\tan\beta_d$.
Care must be taken in defining the heavy Higgs vs. the light Higgs, which also amounts to a definition of $\alpha$.
We provide a comparison of notations in the following.
Reference~\cite{Davoudiasl:2012ag}: $\Phi_1$ couples to the SM fermions, which requires typically small $\tan\beta$ (as $v_1 \sim v$).
$\alpha = 0$ corresponds to the heavy Higgs $\sim$ SM-like Higgs limit.
This paper: $\Phi_2$ couples to the SM fermions, which requires typically large $\tan\beta$ (as $v_2 \sim v$).
$\alpha =\pm \pi/2$ corresponds to the heavy Higgs $\approx$ SM-like Higgs limit.
($\alpha = 0$ corresponds to the light Higgs $\approx$ SM-like Higgs limit.)
The Higgs to gauge boson couplings in Eqs.~\eqref{eq:CHZZ} - \eqref{eq:ChZ'Z'} can be read for Ref.~\cite{Davoudiasl:2012ag} by $\cos\alpha \leftrightarrow \sin\alpha$, $\cos\beta \leftrightarrow \sin\beta$ (or $v_1 \leftrightarrow v_2$), and overall sign flips for ${\cal C}_{hZZ}$, ${\cal C}_{hZZ'}$, ${\cal C}_{hZ'Z'}$.
}
In conventional 2HDMs, one defines $\Phi_2$ to be the doublet that couples to the top quark (or, in the Type I 2HDM, to all fermions).
Following this notation, we assume the $U(1)'$ charges $Q'[\Phi_1] = Q'[\Phi_S] = 1$, $Q'[\Phi_2] = 0$ with the hypercharges $Y[\Phi_1] = Y[\Phi_2] = 1/2$, $Y[\Phi_S] = 0$.
The $U(1)'$ is spontaneously broken when $\Phi_1$ or $\Phi_S$ gets a vacuum expectation value (vev).

The scalar part of the kinetic Lagrangian is given by
\bea
{\cal L}_\text{scalar} &=& | D_\mu \Phi_1 |^2 + | D_\mu \Phi_2 |^2 + | D_\mu \Phi_S |^2 \\
&=& {\cal L}_\text{mass} + {\cal L}_\text{coupling} + \cdots
\eea
where
\beq
D_\mu \Phi_i = \left( \partial_\mu + i g' Y[\Phi_i] \hat B_\mu + i g T_3[\Phi_i] \hat W_{3 \mu} + i g_{Z'} Q'[\Phi_i] \hat Z'^0_\mu \right) \Phi_i
\eeq
and two doublets $\Phi_i = \mat{ \phi_i^+ \\ \frac{1}{\sqrt{2}} (v_i + \phi_i + i \eta_i) }$ with vevs $v_i$ ($i = 1, 2$), and a singlet $\Phi_S = \frac{1}{\sqrt{2}} (v_S + \phi_S + i \eta_S)$ with a vev $v_S$.
The $v_S =0$  limit corresponds to the doublets only case.

The $Z$ and $Z'$ masses, with $v^2 = v_1^2 + v_2^2 \simeq (246 ~\gev)^2$ and $\tan\beta \equiv v_2 / v_1$, $\tan\beta_d \equiv v_S / v_1$ are given by
\beq
{\cal L}_\text{mass} = \frac{1}{2} m_{Z^0}^2 Z^0 Z^0 - \Delta^2 Z^0 Z'^0 + \frac{1}{2} m_{Z'^0}^2 Z'^0 Z'^0 + \cdots
\eeq
where
\bea
m_{Z^0}^2 &=& \frac{1}{4} g_Z^2 v^2 , \\
m_{Z'^0}^2 &=& g_{Z'}^2 (v^2 \cos^2\beta + v_S^2) + \frac{\eps}{\cos\theta_W} g_{Z'} g' v^2 \cos^2\beta + \frac{1}{4} \left(\frac{\eps}{\cos\theta_W}\right)^2 g'^2 v^2 , \label{eq:mZprime0} \\
\Delta^2 &=& \frac{1}{2} g_{Z'} g_Z v^2 \cos^2\beta + \frac{1}{4} \frac{\eps}{\cos\theta_W} g_Z g' v^2 .
\eea
This leads to the $Z$-$Z'$ mixing angle ($\xi$) as
\beq
\tan 2\xi = \frac{2\Delta^2}{m^2_{Z^0}-m^2_{Z^{'0}}}
\eeq
\beq
\mat{Z \\ Z'} = \mat{\cos\xi & -\sin\xi \\ \sin\xi & \cos\xi} \mat{Z^0 \\ Z'^0} \label{eq:massMixing}
\eeq
where $Z$ and $Z'$ are the mass eigenstates.
The $Z$-$Z'$ mixing angle is constrained to be very small ($|\xi| \lsim$ few $\times 10^{-3}$) by precision $Z$ pole measurement at  LEP \cite{Beringer:1900zz}.

We are primarily interested in a very light $Z'$ [$m_{Z'} \lsim {\cal O} (1) ~\gev$].
We will work in the $m_{Z'^0}^2 \ll m_{Z^0}^2$ limit throughout this paper, 
which requires $g_{Z'}^2 (v_1^2 + v_S^2) \ll \frac{1}{4} g_Z^2 v^2$ as $|\eps|$ is very small.
In this limit we have
\bea
m_Z^2 &\simeq& m_{Z^0}^2 = \frac{1}{4} g_Z^2 v^2 , \label{eq:mZ} \\
m_{Z'}^2 &\simeq& m_{Z'^0}^2 - \frac{(\Delta^2)^2}{m_{Z^0}^2} = g_{Z'}^2 ( v^2 \cos^2\beta \sin^2\beta + v_S^2 ) = g_{Z'}^2 v^2 \frac{\cos^2\beta}{\cos^2\beta_d} (1 - \cos^2\beta \cos^2\beta_d) , \label{eq:mZprime} \\
\xi &\simeq& \frac{\Delta^2}{m_{Z^0}^2} = \frac{2 g_{Z'}}{g_Z} \cos^2\beta + \eps \tan\theta_W . \label{eq:xi}
\eea
The $Z'$ approaches the massless limit as $g_{Z'} \to 0$ or $v_1, v_S \to 0$.
With the $\delta$ notation defined in the Appendix , we can use $\cos^2\beta_d \simeq \frac{\delta^2}{1 + \delta^2} \frac{1}{\cos^2\beta}$ (the doublets only limit corresponds to $\delta \tan\beta \simeq 1$), and have
\beq
m_{Z'} \simeq \frac{g_{Z'} v \cos^2\beta}{\delta} .
\eeq

For simplicity, we assume no mixing between the doublets and singlet and allow mixing only between the doublets.
The pure doublets case can be reached if we take $v_S = 0$ (corresponding to $\cos\beta_d = 1$) in Eqs.~\eqref{eq:mZprime0} and \eqref{eq:mZprime}, and it would not change the following couplings in Eqs.~\eqref{eq:CHZZ} - \eqref{eq:ChZ'Z'} as well as the approximations in Eqs.~\eqref{eq:CHZZapprox} - \eqref{eq:ChZ'Z'approx}.

The relevant couplings of vector bosons to heavy Higgs ($H$) and light Higgs ($h$), with no mixing between the doublets and singlet, are
\bea
{\cal L}_\text{coupling} &=& \frac{1}{2} {\cal C}_{HZZ} HZZ + {\cal C}_{HZZ'} HZZ'+ \frac{1}{2} {\cal C}_{HZ'Z'} HZ'Z' \nn \\
&+& \frac{1}{2} {\cal C}_{hZZ} hZZ + {\cal C}_{hZZ'} hZZ'+ \frac{1}{2} {\cal C}_{hZ'Z'} hZ'Z' + \cdots
\eea
where
\bea
{\cal C}_{HZZ}   &=& {\cal C}_{HZZ}^\text{SM} \left( \cos\beta \cos\alpha \left[ \cos\xi + (2 g_{Z'}/g_Z + \eps \tan\theta_W) \sin\xi \right]^2 + \sin\beta \sin\alpha \left[ \cos\xi + \eps \tan\theta_W \sin\xi\right]^2 \right) \label{eq:CHZZ} \\
{\cal C}_{HZZ'}  &=& {\cal C}_{HZZ}^\text{SM}  \left( \cos\beta \cos\alpha\left[ \cos\xi + (2 g_{Z'}/g_Z + \eps \tan\theta_W) \sin\xi \right] \left[ \sin\xi - (2 g_{Z'}/g_Z + \eps \tan\theta_W) \cos\xi \right] \right. \nn \\
&& \hspace{10mm} \left. + \sin\beta \sin\alpha \left[ \cos\xi + \eps \tan\theta_W \sin\xi\right] \left[ \sin\xi - \eps \tan\theta_W \cos\xi \right] \right) \\
{\cal C}_{HZ'Z'} &=& {\cal C}_{HZZ}^\text{SM} \left( \cos\beta \cos\alpha \left[ \sin\xi - (2 g_{Z'}/g_Z + \eps \tan\theta_W) \cos\xi \right]^2 + \sin\beta \sin\alpha \left[ \sin\xi - \eps \tan\theta_W \cos\xi\right]^2 \right) 
\eea
and
\bea
{\cal C}_{hZZ}   &=& {\cal C}_{HZZ}^\text{SM} \left( \sin\beta \cos\alpha  \left[ \cos\xi + \eps \tan\theta_W \sin\xi\right]^2 - \cos\beta \sin\alpha \left[ \cos\xi + (2 g_{Z'}/g_Z + \eps \tan\theta_W) \sin\xi \right]^2 \right) \\
{\cal C}_{hZZ'}  &=& {\cal C}_{HZZ}^\text{SM} \left( \sin\beta \cos\alpha  \left[ \cos\xi + \eps \tan\theta_W \sin\xi\right] \left[ \sin\xi - \eps \tan\theta_W \cos\xi \right]  \right. \nn \\
&& \hspace{10mm} \left. - \cos\beta \sin\alpha \left[ \cos\xi + (2 g_{Z'}/g_Z + \eps \tan\theta_W) \sin\xi \right] \left[ \sin\xi - (2 g_{Z'}/g_Z + \eps \tan\theta_W) \cos\xi \right] \right) \\
{\cal C}_{hZ'Z'} &=& {\cal C}_{HZZ}^\text{SM} \left( \sin\beta \cos\alpha \left[ \sin\xi - \eps \tan\theta_W \cos\xi \right]^2 - \cos\beta \sin\alpha \left[ \sin\xi - (2 g_{Z'}/g_Z + \eps \tan\theta_W) \cos\xi \right]^2 \right) \label{eq:ChZ'Z'}
\eea
with the SM Higgs-$Z$-$Z$ coupling ${\cal C}_{HZZ}^\text{SM} \equiv \frac{1}{2} g_Z^2 v$.
Note that, unlike Ref.~\cite{Davoudiasl:2012ag}, we include couplings with a general mixing angle $\alpha$.

Since $|\xi|$, $|\eps| \ll 1$, we can get approximations, using Eq.~\eqref{eq:xi},
\bea
{\cal C}_{HZZ}   &\simeq& {\cal C}_{HZZ}^\text{SM} \cos(\beta - \alpha) \label{eq:CHZZapprox} \\
{\cal C}_{HZZ'}  &\simeq& - {\cal C}_{HZZ}^\text{SM} (2 g_{Z'} / g_Z) \cos\beta \sin\beta \sin(\beta - \alpha) \label{eq:CHZZ'approx}\\
{\cal C}_{HZ'Z'} &\simeq& {\cal C}_{HZZ}^\text{SM} (2 g_{Z'} / g_Z)^2 \cos\beta \sin\beta \left( \cos^3\beta \sin\alpha + \sin^3\beta \cos\alpha \right) \label{eq:CHZ'Z'approx}
\eea
and
\bea
{\cal C}_{hZZ}   &\simeq& {\cal C}_{HZZ}^\text{SM} \sin(\beta - \alpha) \label{eq:ChZZapprox} \\
{\cal C}_{hZZ'}  &\simeq& {\cal C}_{HZZ}^\text{SM} (2 g_{Z'} / g_Z) \cos\beta \sin\beta \cos(\beta - \alpha) \label{eq:ChZZ'approx} \\
{\cal C}_{hZ'Z'} &\simeq& {\cal C}_{HZZ}^\text{SM} (2 g_{Z'} / g_Z)^2 \cos\beta \sin\beta \left( \cos^3\beta \cos\alpha - \sin^3\beta \sin\alpha \right) \label{eq:ChZ'Z'approx}
\eea
giving the expected 2HDM couplings of the two neutral Higgs bosons to the $Z$ in Eqs.~\eqref{eq:CHZZapprox} and \eqref{eq:ChZZapprox}.
In the $\alpha = \pi/2$ limit, we can reproduce the relation shown in Ref.~\cite{Davoudiasl:2012ag}, i.e. $\eps_Z \simeq {\cal C}_{HZZ'} / {\cal C}_{HZZ} \simeq {\cal C}_{HZ'Z'} / {\cal C}_{HZZ'}$.

The scalar potential is
\bea
V &=& V_1 + V_2 \\
V_1 &=& m_{11}^2 \Phi_1^\dagger \Phi_1 + m_{22}^2 \Phi_2^\dagger \Phi_2 + \frac{\lambda_1}{2} ( \Phi_1^\dagger \Phi_1 )^2 + \frac{\lambda_2}{2} ( \Phi_2^\dagger \Phi_2 )^2 + \lambda_3 ( \Phi_1^\dagger \Phi_1 ) ( \Phi_2^\dagger \Phi_2 ) + \lambda_4 ( \Phi_1^\dagger \Phi_2 ) ( \Phi_2^\dagger \Phi_1 ) \\
V_2 &=& m_{33}^2 \Phi_S^\dagger \Phi_S + \frac{\lambda_6}{2} ( \Phi_S^\dagger \Phi_S )^2
\eea
with all coefficients real.
Interestingly, the terms $m_{12}^2 ( \Phi_1^\dagger \Phi_2 + \Phi_2^\dagger \Phi_1 )$ and $\frac{\lambda_5}{2} [(\Phi_1^\dagger \Phi_2)^2 + (\Phi_2^\dagger \Phi_1)^2]$ whose coefficients are generally complex are forbidden by the $U(1)'$ gauge symmetry.
We assume the terms $\lambda_7 ( \Phi_1^\dagger \Phi_1 ) ( \Phi_S^\dagger \Phi_S ) + \lambda_8 ( \Phi_2^\dagger \Phi_2 ) ( \Phi_S^\dagger \Phi_S )$ are negligible (no mixing between the doublets and singlet).

The charged Higgs mass is given by
\beq
m_{H^\pm}^2 = -\frac{\lambda_4}{2} v^2
\eeq
which requires $\lambda_4 < 0$.

The Higgs singlet mass is given by
\beq
m_S^2 = \lambda_6 v_S^2
\eeq
and the neutral doublet Higgs mass-squared matrix is given by
\beq
M_\text{Higgs}^2 = \mat{\lambda_1 v_1^2 & (\lambda_3 + \lambda_4) v_1 v_2 \\ (\lambda_3 + \lambda_4) v_1 v_2 & \lambda_2 v_2^2} . \label{eq:HiggsMatrix}
\eeq
The mass eigenstates of the doublets are $H$ and $h$ (with $m_H \ge m_h$) with
\bea
m_H^2 = \frac{1}{2} \left( \lambda_1 v_1^2 + \lambda_2 v_2^2 + \sqrt{ (\lambda_1 v_1^2 - \lambda_2 v_2^2)^2 + 4 (\lambda_3 + \lambda_4)^2 v_1^2 v_2^2} \right) \\
m_h^2 = \frac{1}{2} \left( \lambda_1 v_1^2 + \lambda_2 v_2^2 - \sqrt{ (\lambda_1 v_1^2 - \lambda_2 v_2^2)^2 + 4 (\lambda_3 + \lambda_4)^2 v_1^2 v_2^2} \right)
\eea
and the $H$-$h$ mixing angle is given by
\beq
\tan 2\alpha = \frac{2(\lambda_3 + \lambda_4)v_1 v_2}{\lambda_1 v_1^2 - \lambda_2 v_2^2}
\eeq
\beq
\mat{H \\ h} = \mat{\cos\alpha & \sin\alpha \\ -\sin\alpha & \cos\alpha} \mat{\phi_1 \\ \phi_2} .
\eeq

Since only $\Phi_2$ couples to the SM fermions in this model, the $\sin\alpha \approx \pm 1$ limit provides the heavier Higgs ($H$) as the SM-like Higgs ($H \sim \phi_2$, $h \sim \phi_1$ up to a sign), which is the case of interest in the subsequent sections in this paper.
The other limit, $\sin\alpha \approx 0$, would have provided the light Higgs ($h$) as the SM-like Higgs ($H \sim \phi_1$, $h \sim \phi_2$ up to a sign).
Note that the SM-like limit for the $H$ is achieved for $\beta = \alpha$ instead of $\beta = \alpha + \pi / 2$ which would be true when the light Higgs $h$ is the SM-like one.

The relative coupling of the Higgses to the SM fermions and $W$ boson in the Dark 2HDM is the same as the Type I 2HDM as following.
\bea
Htt,~ Hbb,~ H\tau\tau:~ \frac{\sin\alpha}{\sin\beta}, && HWW:~ \cos(\beta-\alpha) ~~
\label{eq:Hcoupling}
\\
htt,~ hbb,~ h\tau\tau:~ \frac{\cos\alpha}{\sin\beta}, && hWW:~ \sin(\beta-\alpha) ~~
\label{eq:hcoupling}
\eea
The relative coupling of $HZZ$ ($hZZ$) is the same as that of the $HWW$ ($hWW$) to a good approximation since the $Z$ mixing is very small [Eq.~\eqref{eq:CHZZapprox}, Eq.~\eqref{eq:ChZZapprox}].

\section{\boldmath LEP and LHC Constraints on the Neutral Higgs Bosons, $h$ and $H$}
\label{sec:constraints}
\begin{figure*}[bt]
\begin{center}
\includegraphics[height=0.325\textwidth]{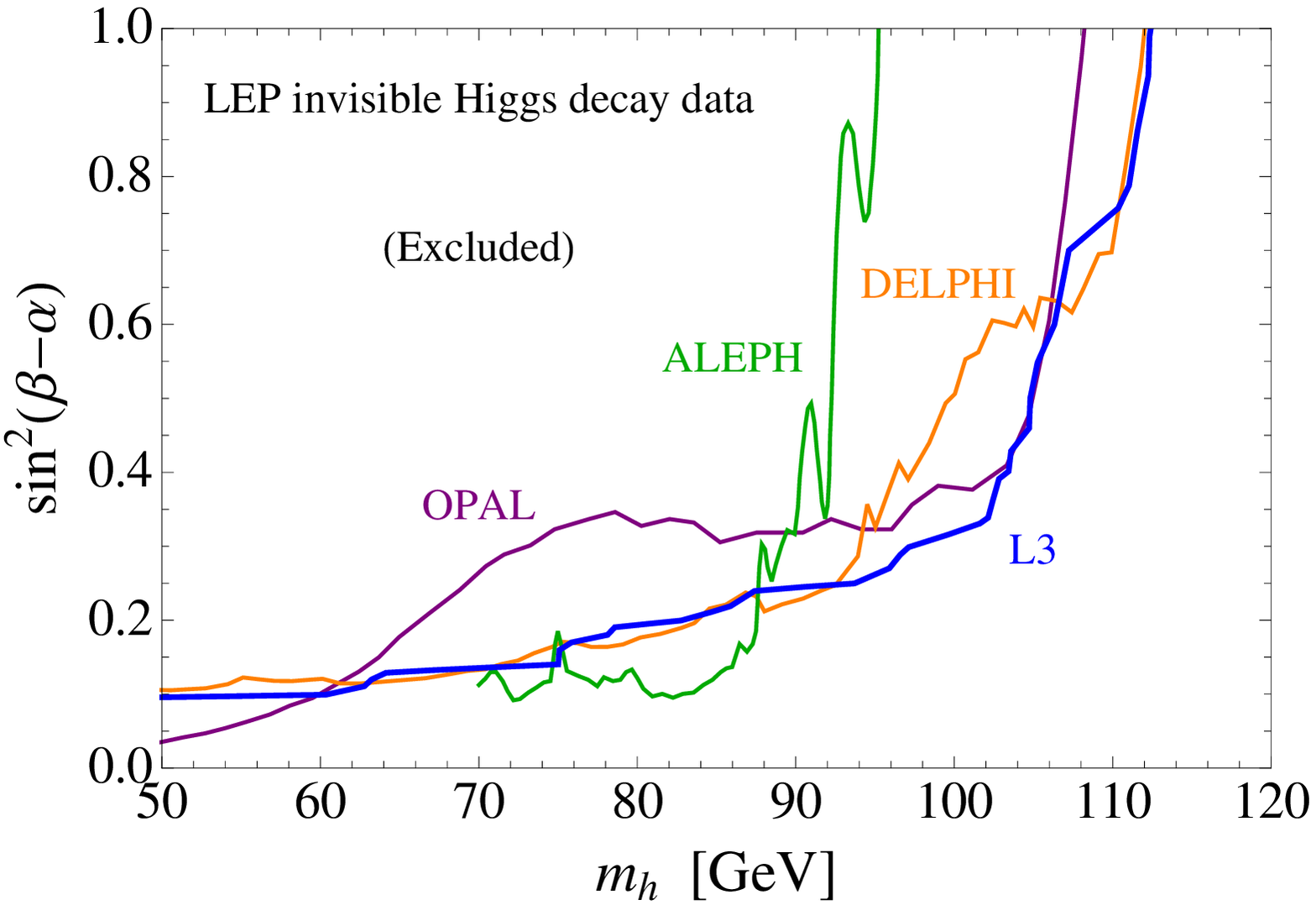} ~~~~~
\includegraphics[height=0.325\textwidth]{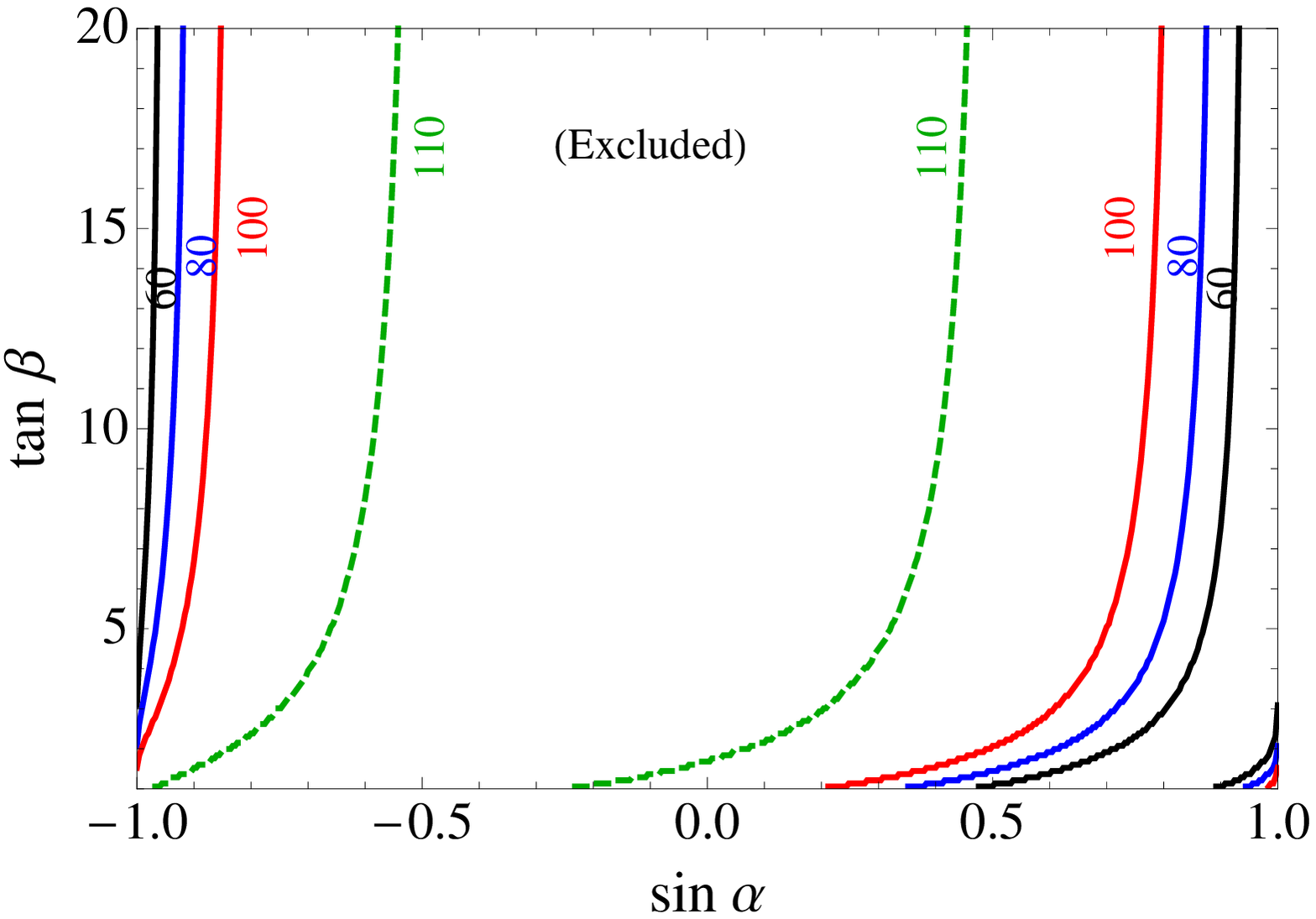}\\
~~~~~~ (a) ~~~~~~~~~~~~~~~~~~~~~~~~~~~~~~~~~~~~~~~~~~~~~~~~~~~~~~~~~~~~~~~~~~~~~~~~~~~~~~~ (b)
\end{center}
\caption{
(a) LEP constraints on the invisible Higgs decay $\sigma(Zh) \br(h \to \text{inv}) / \sigma(ZH_\text{SM})$ at $95 \%$ CL.
It is the same as $\sin^2(\beta-\alpha)$ when small $Z$ mixing is neglected.
(b) LEP invisibly decaying Higgs bounds (L3) on $\alpha$ and $\beta$ in all 2HDMs including the Dark 2HDM (when the $Z$ mixing is neglected).
The parameter-space within the bands (central part) are excluded, for a given light Higgs mass $m_h = 60 ~\gev$ (black/darkest), $80 ~\gev$ (blue/dark), $100 ~\gev$ (red/gray), and $110 ~\gev$ (green/dashed).
The dashed curve (for $110 ~\gev$) shows how quickly the LEP bounds converges for $m_h \gsim 105 ~\gev$.
}
\label{fig:LEP}
\end{figure*}

In this section, we consider various constraints/predictions on the neutral Higgs bosons in the Dark 2HDM.
They include
(i) LEP limits on associated production of the light Higgs ($Z^* \to Z h$),
(ii) LEP limits on the marginal $Z$ width for $Z \to h Z'$ decay,
(iii) LHC measured branching ratios in various modes ($H \to b \bar b$, $\tau^+\tau^-$, $WW$, $ZZ$) including the most precisely measured $H \to \gamma\gamma$.

We will focus on the case in which the mass of the light Higgs is $m_H / 2 \lsim m_h \lsim m_Z$, i.e. roughly, $m_h \simeq 60 - 90 ~\gev$ range.
If it is lighter than $m_H / 2$, then the fact that $H \to h h$ decays do not dominate restricts $\sin\alpha$ to be infinitesimally close to $\sin\alpha = 1$ \cite{Ferreira:2012my}.
If it is heavier than $m_Z$, then the $h \to Z Z'$ channel opens up although our results may not change qualitatively.

\subsection{LEP bounds from associated production of $h$}
A SM Higgs of less than $114$ GeV was ruled out long ago by LEP experiments looking for associated production with a $Z$.
In 2HDMs, however, the $hWW$ and $hZZ$ (neglecting small $Z$ mixing) couplings are suppressed by $\sin(\beta-\alpha)$, possibly making the $h$ quite gaugephobic.
LEP published upper bounds on this factor as a function of the Higgs mass, assuming the Higgs would decay into $b$ quarks or $\tau$'s \cite{Barate:2003sz,Schael:2006cr,Ferreira:2009jb}.
However, as we will discuss in Sec.~\ref{sec:lightHiggs}, $h$ can mainly decay into $Z'Z'$ in the Dark 2HDM.
Since the LEP searches did not cover fermions from a very light $Z'$ [$m_{Z'} \lsim {\cal O}(1) ~\gev$], one must look at LEP bounds on invisible Higgs decays.
These bounds at $95 \%$ CL from ALEPH \cite{Barate:1999uc}, DELPHI \cite{Abdallah:2003ry}, L3 \cite{Achard:2004cf}, OPAL \cite{Abbiendi:2007ac} collaborations are given in Fig.~\ref{fig:LEP}~(a).
There are also updated ALEPH results, which covers the Higgs mass from $95 ~\gev$ \cite{Heister:2001kr}, but this is beyond our  range of interest.  For simplicity, we will take only L3 data, which is roughly close to average values of the LEP data in the $m_h$ range we consider in this paper ($m_h \simeq 60 - 90 ~\gev$).

For a given $h$ mass, this bound is given as a band in $\tan\beta - \sin\alpha$ plane in Fig.~\ref{fig:LEP}~(b).
The apparent asymmetry between the $\sin\alpha \sim -1$ and $\sin\alpha \sim 1$ regions originates from $\tan\beta > 0$, which makes the region near $\sin\alpha \sim 1$ (or $\alpha \sim \pi / 2$) preferable .
As the $h$ mass increases from around $105 ~\gev$, the bands quickly converge as the LEP excluded region vanishes [Fig.~\ref{fig:LEP}~(a)].
This is illustrated as a sudden departure of $m_h = 110 ~\gev$ curve from the $m_h = 100 ~\gev$ curve in Fig.~\ref{fig:LEP} (b).

\subsection{LEP bounds from $Z \to h Z'$}
\label{sec:3B}
The $h$ can be produced directly from $Z$ decays into an $h$ and a $Z'$, if kinematically allowed.
It will appear as a contribution to the invisible $Z$ width.

For a sufficiently light $Z'$, using Eq.~\eqref{eq:ChZZ'approx}, we have
\bea
\Gamma (Z \to h Z') &\simeq& ({\cal C}_{hZZ'})^2 \frac{m_Z}{64\pi m_{Z'}^2} \left( 1 - \frac{m_h^2}{m_Z^2} \right)^3 \\
&\simeq& \frac{g_Z^2 m_Z}{64 \pi} \left( \delta \tan\beta \right)^2 \cos^2 (\beta-\alpha) \left( 1 - \frac{m_h^2}{m_Z^2} \right)^3 .
\eea
As was discussed in Ref.~\cite{Davoudiasl:2012ag}, the boosted $Z'$ shows the divergent nature as $m_{Z'} \to 0$ due to the enhancement from the longitudinal component of the $Z'$.

\begin{figure*}[bt]
\begin{center}
\includegraphics[height=0.325\textwidth]{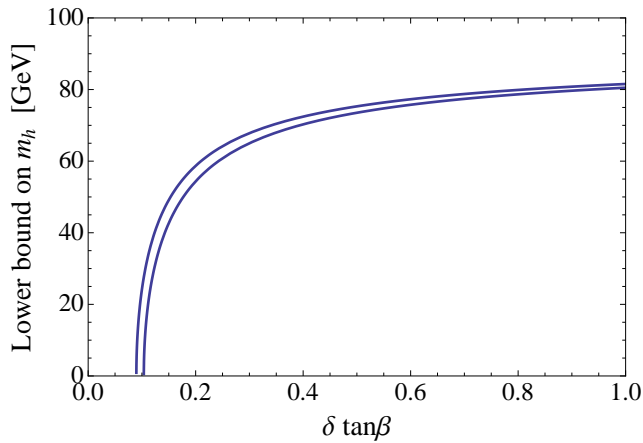}
\end{center}
\caption{
Lower bound on the light Higgs mass from the $Z \to h Z'$ contribution to the invisible width. Upper (Lower) curve is for the extreme value of $\cos^2(\beta - \alpha) = 1.0 ~ (0.75)$.
The doublets only (no singlet) case corresponds to $\delta\tan\beta \simeq 1$ point.
}
\label{fig:mhBound}
\end{figure*}

The limit on the new physics contribution to the undetected width of the $Z$, which can be obtained by subtracting the SM prediction from the measured width, is $2 ~\mev$ (95\% CL) from  LEP data \cite{Carena:2003aj}.
For $h$ masses below the $Z$ mass, we can see from Fig.~\ref{fig:LEP}~(a) that $\cos^2(\beta - \alpha)$ is always greater than $0.75$ for L3.
The lower bound on $m_h$ is given as a function of $\delta\tan\beta$ in Fig.~\ref{fig:mhBound} for two extreme values of $\cos^2(\beta - \alpha)$.

In the doublets only case ($\delta\tan\beta \simeq 1$), we would have had $m_h \gsim 80 ~\gev$.
Since the $m_h$ and $m_H ~(\simeq 125 ~\gev)$ would then be of similar size, $v_1$ and $v_2$ would be also of similar size, unless some of the $\lambda_i$ values in the Higgs mass-squared matrix in Eq.~\eqref{eq:HiggsMatrix} are larger than $4\pi$, violating perturbativity.   The requirement of perturbativity then 
gives a limit of $\tan\beta \lsim 10$ (or $\delta \gsim 0.1$).
As discussed in the Appendix, experimental constraints provide a tight bound on $\delta$ in the doublets only case  which conflicts with this limit.
With a presence of the Higgs singlet, $\delta \tan\beta$ can be lower than $1$, and the $m_h$ bound gets relaxed, as Fig.~\ref{fig:mhBound} shows.
For $\delta \tan\beta \lsim 0.1$, there is essentially no bound on $m_h$ from the LEP invisible decay width.
Thus, we find that the doublets only case is unacceptable in the scenario we consider.

We see that for sufficiently small $\delta \tan\beta$, one can have $m_h < 62.5 ~ \gev$, leading to the possibility of $H\to hh$.   As shown in Ref. \cite{Ferreira:2012my}, this decay would unacceptably dominate $H$ decays unless $\sin\alpha$ was infinitesimally close to $1$.   We will not consider this possibility in this paper.

\subsection{\boldmath LHC bounds from the $125 ~\gev$ SM-like Higgs}
\label{sec:125bounds}
The discovery of a Higgs boson, $H$, at $125$ GeV with roughly  SM branching ratios will imply stringent constraints on additional decays of the $H$.
In particular, the new decays $H \to ZZ'$ and $H \to Z'Z'$ should be sufficiently small.

For a sufficiently light $Z'$, using Eqs.~\eqref{eq:CHZZ'approx} and \eqref{eq:CHZ'Z'approx}, we have
\beq
\Gamma (H \to Z Z') \simeq \frac{g_Z^2}{64 \pi} \frac{(m_H^2 - m_Z^2)^3}{m_H^3 m_Z^2} \left( \delta \tan\beta \right)^2 \sin^2 (\beta - \alpha)
\eeq
and
\beq
\Gamma (H \to Z' Z') \simeq \frac{g_Z^2}{128 \pi} \frac{m_H^3}{m_Z^2} \left( \delta \tan\beta \right)^4 \left( \frac{\cos^3\beta \sin\alpha + \sin^3\beta \cos\alpha}{\cos\beta \sin\beta} \right)^2 .
\eeq

\begin{figure*}[t]
\begin{center}
\includegraphics[height=0.325\textwidth]{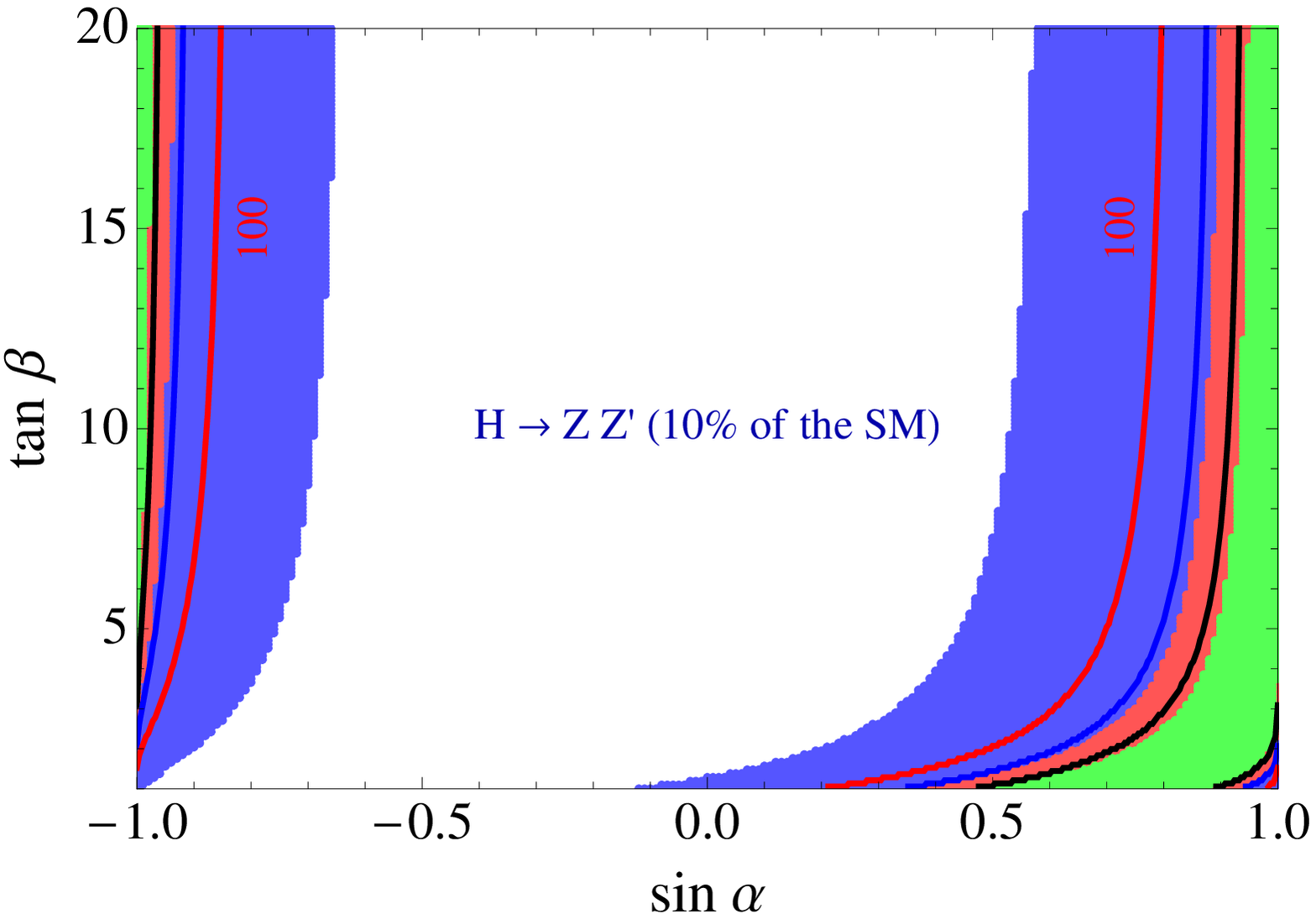} ~~~~~
\includegraphics[height=0.325\textwidth]{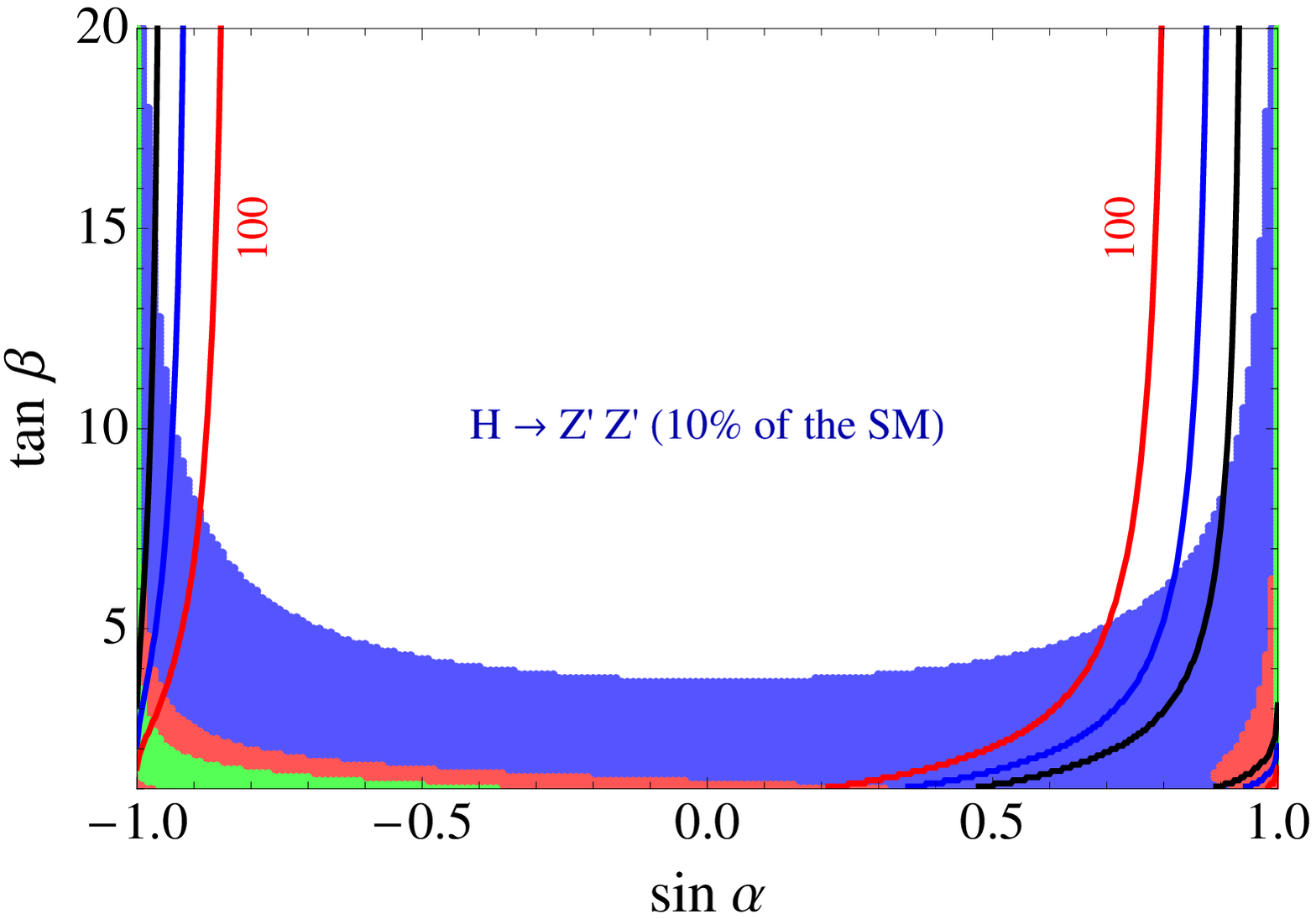}\\
~~~~~~ (a) ~~~~~~~~~~~~~~~~~~~~~~~~~~~~~~~~~~~~~~~~~~~~~~~~~~~~~~~~~~~~~~~~~~~~~~~~~~~~~~~ (b)
\end{center}
\caption{
Decays of the $125 ~\gev$ Higgs to $Z$ and $Z'$, with (a) representing $H \to Z Z'$ and (b) representing $H \to Z' Z'$.
The colored bands are the areas where $H \to Z Z'$, $H \to Z' Z'$ are within $10 \%$ of the SM width.
$\delta \tan\beta = 0.1$ (blue/dark gray band), $0.2$ (red/medium gray band), $0.3$ (green/light gray band) are illustrated.
The smaller $\delta \tan\beta$ region covers the region of larger $\delta \tan\beta$, i.e. the red/medium gray (green/light gray) band is within the blue/dark gray (red/medium gray) band.
}
\label{fig:Darkdecays}
\end{figure*}

In Fig.~\ref{fig:Darkdecays}, we show the region in which their partial decay widths are within $10 \%$ of the SM total decay widths ($4.1 ~\mev$ for the $125 ~\gev$ Higgs), illustrating the sensitive parameter-space for each decay for a few values of $\delta \tan\beta = 0.1$, $0.2$, $0.3$.  In the region between the colored bands, the partial decay widths are more than $10 \%$ of the SM total decay width.
A very light $Z'$ at the LHC could be found by a narrow resonance search after appropriate cuts as studied in Refs.~\cite{Davoudiasl:2012ag,BNL}.
Constraints on the model from these decay modes need a dedicated LHC data analysis.

\begin{figure*}[tb]
\begin{center}
\includegraphics[height=0.325\textwidth]{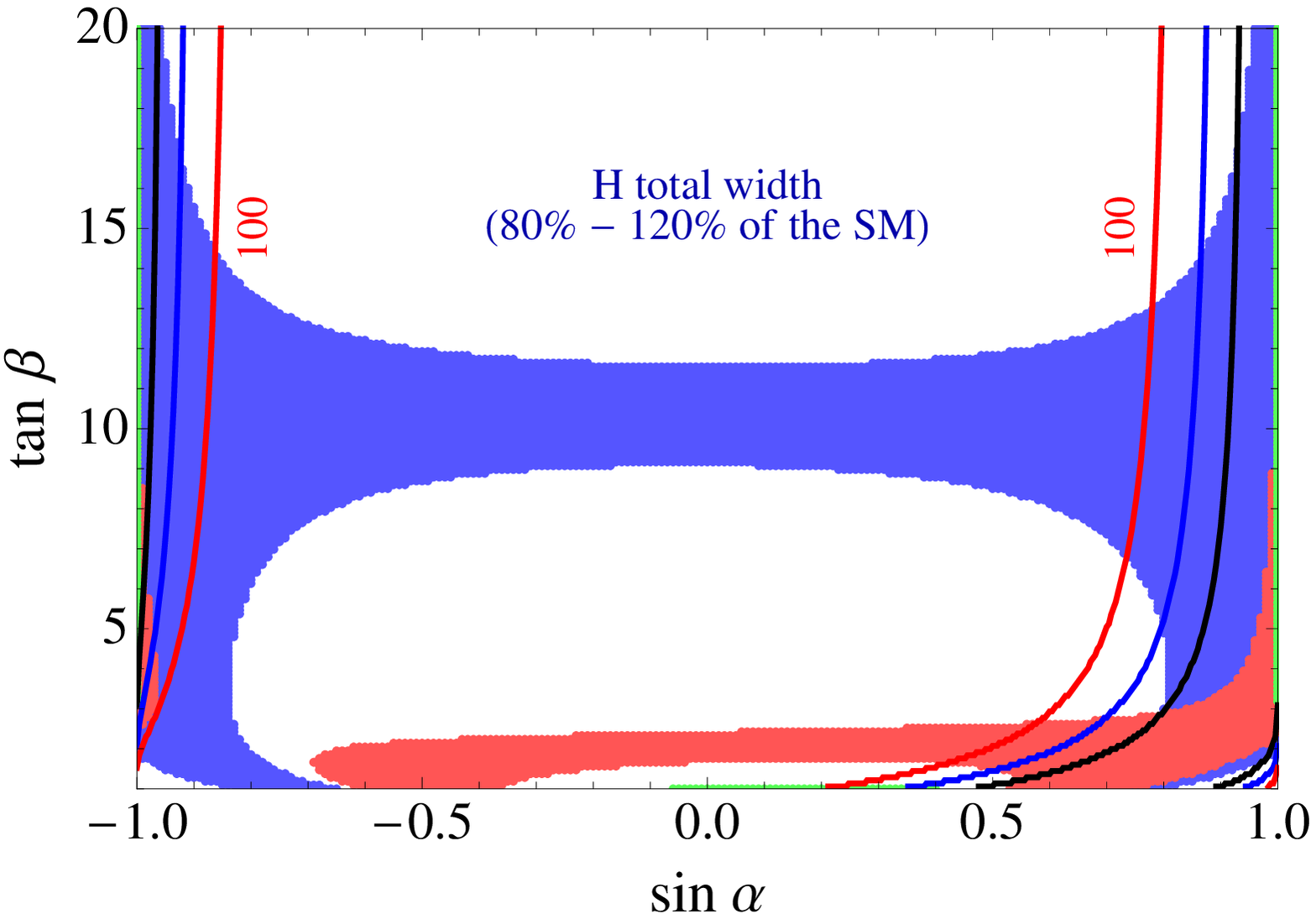} ~~~~~
\includegraphics[height=0.325\textwidth]{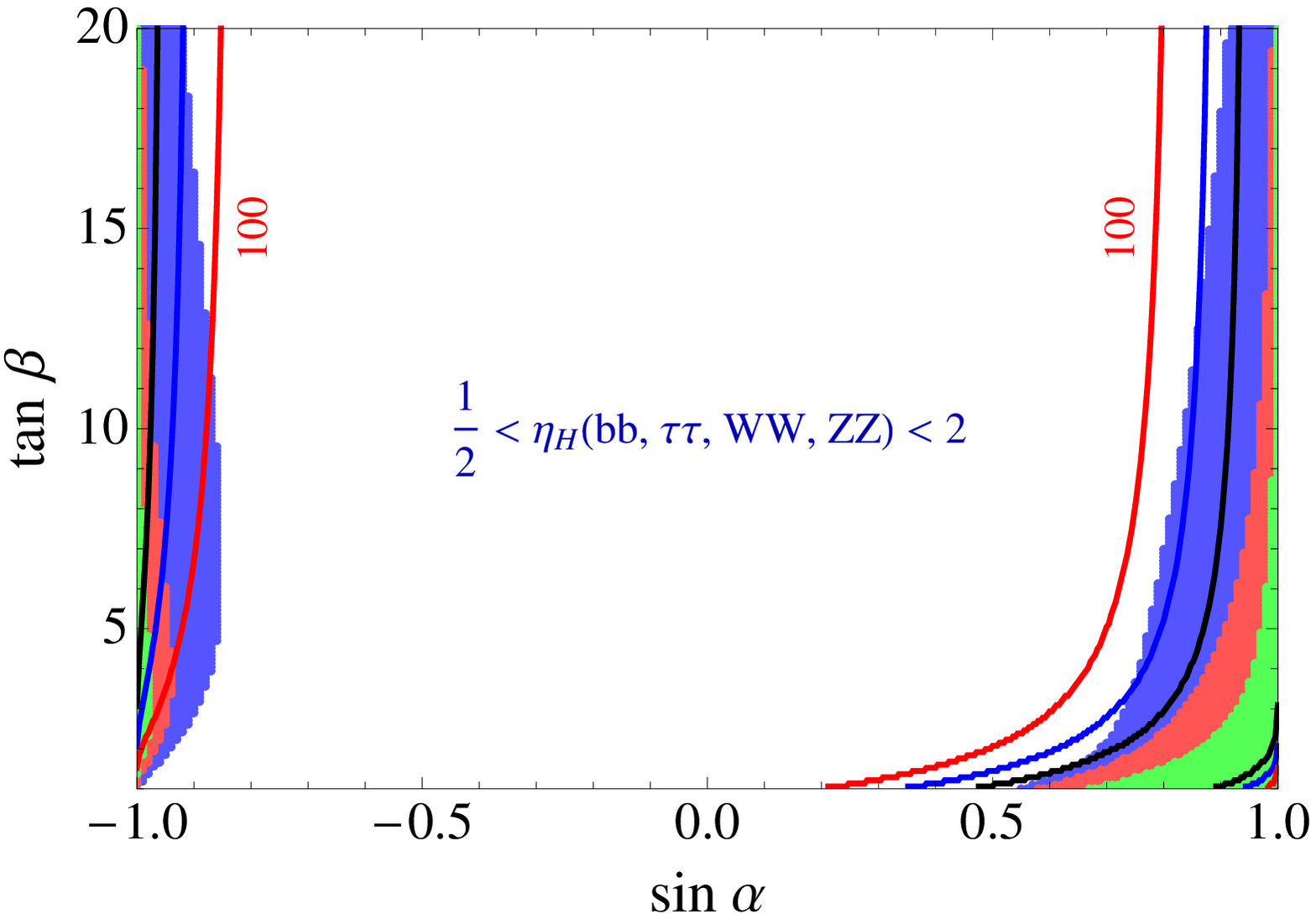}\\
~~~~~~ (a) ~~~~~~~~~~~~~~~~~~~~~~~~~~~~~~~~~~~~~~~~~~~~~~~~~~~~~~~~~~~~~~~~~~~~~~~~~~~~~~~ (b)
\end{center}
\caption{
(a) Region where the total $\Gamma_H$ is within $20 \%$ of the the SM prediction.
(b) Region consistent with the SM predictions within a factor of two in all $b \bar b$, $\tau^+\tau^-$, $WW$, $ZZ$ channels.
Choices of $\delta \tan\beta$ and color code are the same as Fig.~\ref{fig:Darkdecays}.
}
\label{fig:MoreConstraints}
\end{figure*}

Figure~\ref{fig:MoreConstraints} (a) shows the parameter-space where the total decay width of the heavy Higgs $H$ is close to the SM width (within $20 \%$).
We have used a modified version of the code of Ref. \cite{Ferreira:2012my} for some of the plots in this section.
It shows that for a relatively large $\delta \tan\beta$, the $\Gamma_H$ would be quite different from the SM prediction except for a very narrow region of parameter-space.

The scenario that the heavy Higgs is the SM-like Higgs in ordinary 2HDMs were discussed in Ref. \cite{Ferreira:2012my}.
We follow their parametrization to measure the event rate ratio compared to the SM predictions.
For the $H \to \gamma\gamma$,
\beq
\eta_H (\gamma\gamma) = \left(\frac{\sin\alpha}{\sin\beta}\right)^2 \frac{\br^\text{Dark}_\text{2HDM}(H\to \gamma\gamma)}{\br_\text{SM}(H\to \gamma\gamma)}
\eeq
with $\br_\text{SM}(H\to \gamma\gamma)$ being our calculated value, instead of the precise value of the SM \cite{Denner:2011mq}.
Similarly, $\eta_H ( b \bar b )$, etc. are defined for other decay modes.
The prefactor comes from the relative Yukawa coupling of the $H$ and SM quarks [Eq.~\eqref{eq:Hcoupling}], which is relevant for the dominant gluon fusion $g g \to H$ through quark loops.

As noted in Ref.~\cite{Ferreira:2012my}, the number of events (production cross section times branching ratio) of the $H$ into many modes can vary up a factor of two, given currently available data.
Figure~\ref{fig:MoreConstraints} (b) shows the parameter-space in the Dark 2HDM where the branching ratios of the $b \bar b$, $\tau^+ \tau^-$, $WW$ and $ZZ$ are consistent with the SM prediction up to a factor of two ($\frac{1}{2} < \eta_H < 2$).
In the region between the bands, at least one of these event rate ratios is more than a factor of two different from the SM prediction.

\begin{figure*}[bt]
\begin{center}
\includegraphics[height=0.325\textwidth]{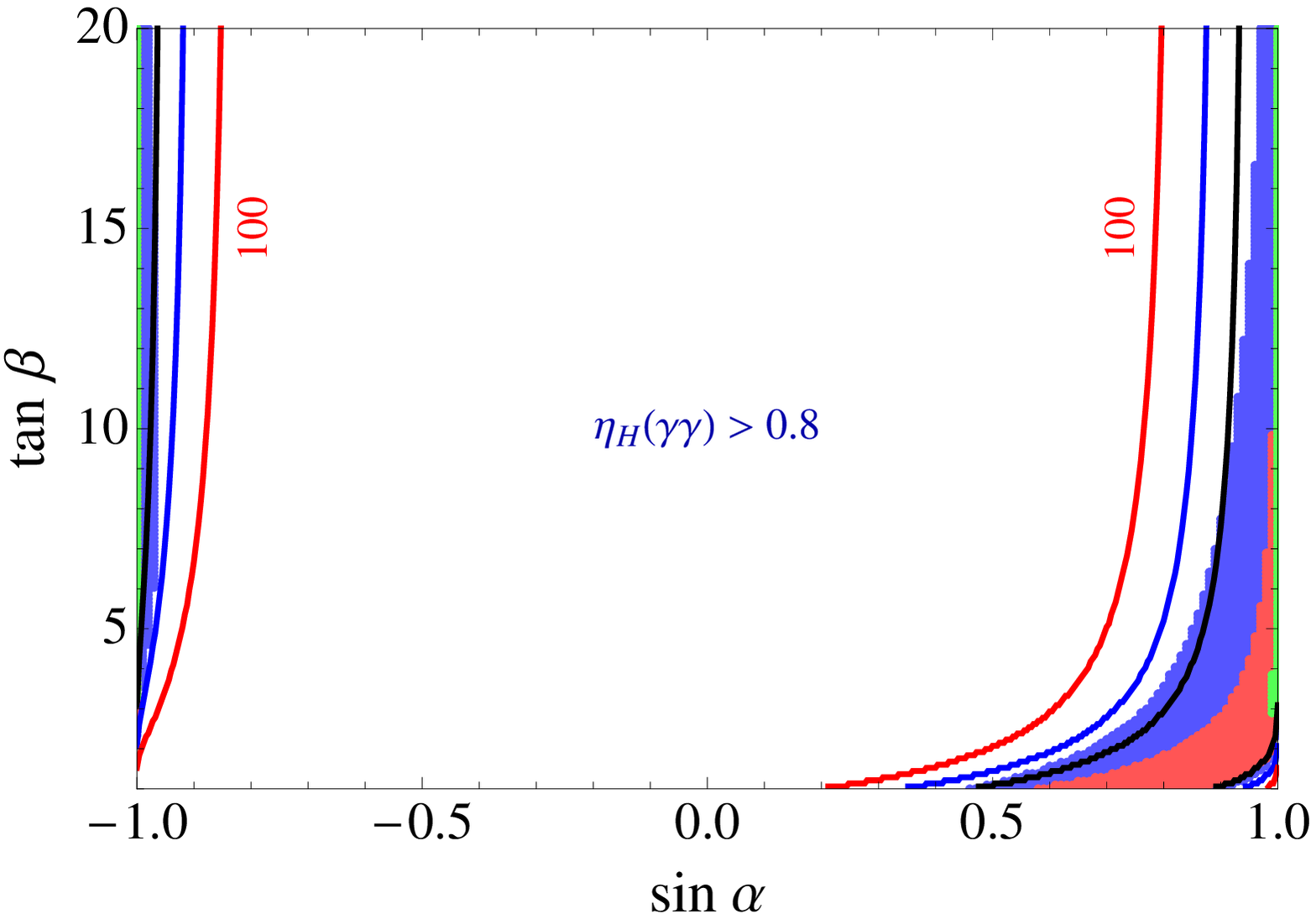} ~~~~~
\includegraphics[height=0.325\textwidth]{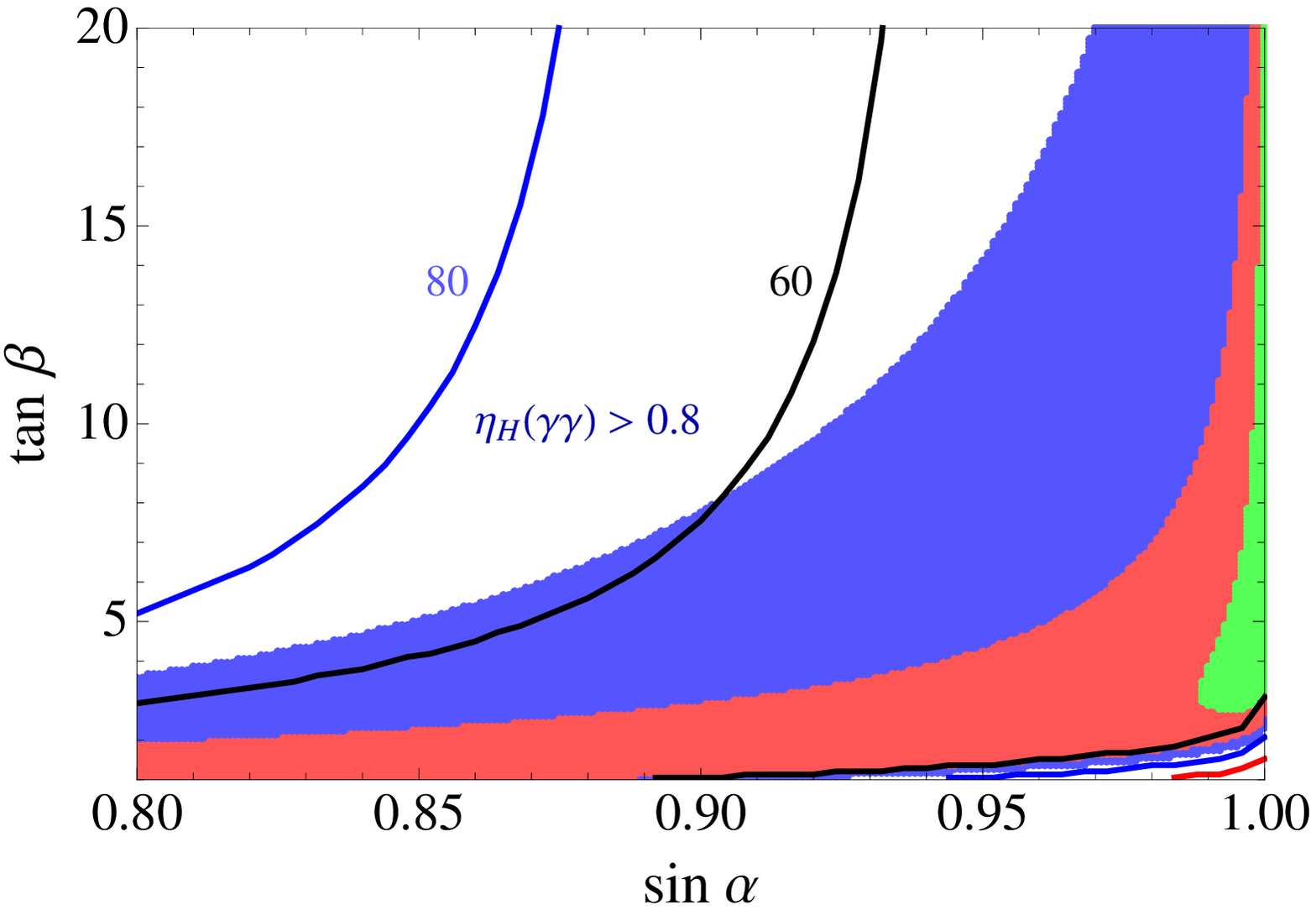}\\
~~~~~~ (a) ~~~~~~~~~~~~~~~~~~~~~~~~~~~~~~~~~~~~~~~~~~~~~~~~~~~~~~~~~~~~~~~~~~~~~~~~~~~~~~~ (b)
\end{center}
\caption{
(a) Region where the diphoton rate ratio is $\eta_H (\gamma\gamma) > 0.8$.
(b) Close-up version of far-right part of the (a).
Choices of $\delta \tan\beta$ and color code are the same as Fig.~\ref{fig:Darkdecays}.
}
\label{fig:diphoton}
\end{figure*}

The $\gamma\gamma$ mode is the decay mode that drove the recent discovery of the $125 ~\gev$ Higgs boson at the LHC.
We plot the allowed region corresponding to $\eta_H (\gamma\gamma) > 0.8$ in Fig.~\ref{fig:diphoton}.
The exact $\eta_H (\gamma\gamma) = 1$ is not achievable with given choices of $\delta\tan\beta$ ($= 0.1$, $0.2$, $0.3$), and requires much smaller values to reach this limit.
In the ordinary Type I model, it is well known that one can not substantially exceed $\eta_H (\gamma\gamma) = 1$.

One can expand the allowed parameter-space considerably by going to smaller $\delta\tan\beta$ values.
We also note the parameter-space of the diphoton constraint in Fig.~\ref{fig:diphoton} is covered, except for a very little portion at the edge, by the parameter-space of less stringent other modes in Fig.~\ref{fig:MoreConstraints} (b).

The effect of the charged Higgs that can contribute to diphoton signals may depend on the parameters of the scalar potential, and it is ignored (which is valid when the charged Higgs is heavy enough).
We will consider relatively light charged Higgs scenario later in this paper though, which can change the diphoton rate.
It is worth mentioning that $\eta_H (\gamma\gamma) \approx 1$ may not be a decisive constraint as additional vector-like leptons can increase the diphoton decay rate although they can also alter other decays \cite{Davoudiasl:2012ig,BNL}.

\section{\boldmath Decays and detection of light Higgs, $h$}
\label{sec:lightHiggs}
The most interesting decay of the $h$ would be $h \to Z'Z'$.
(For simplicity, we consider only the $m_h \lsim m_Z$ case to avoid $h \to Z Z'$ decay.)

For a sufficiently light $Z'$, using Eq.~\eqref{eq:ChZ'Z'approx}, we have
\beq
\Gamma (h \to Z' Z') \simeq \frac{g_Z^2}{128 \pi} \frac{m_h^3}{m_Z^2} \left( \delta \tan\beta \right)^4 \left(\frac{\cos^3\beta \cos\alpha - \sin^3\beta \sin\alpha}{\cos\beta \sin\beta} \right)^2 .
\eeq
We can compare this to the typically dominant decay channel, $b \bar b$,
\beq
\Gamma (h \to b\bar{b}) \simeq \frac{3m^2_bm_h}{8\pi v^2} \left( \frac{\cos\alpha}{\sin\beta} \right)^2
\eeq
which gives
\beq
\frac{\Gamma (h \to b\bar{b})}{\Gamma (h \to Z' Z')} = \frac{12 m_b^2}{m_h^2} \frac{1}{\left(\delta \tan\beta\right)^4} \left( \frac{\cos\beta \sin\beta}{\cos^3\beta \cos\alpha - \sin^3\beta \sin\alpha} \right)^2   \left( \frac{\cos\alpha}{\sin\beta} \right)^2.
\label{eq:lightHiggsRatio}
\eeq

\begin{figure*}[tb]
\begin{center}
\includegraphics[height=0.325\textwidth]{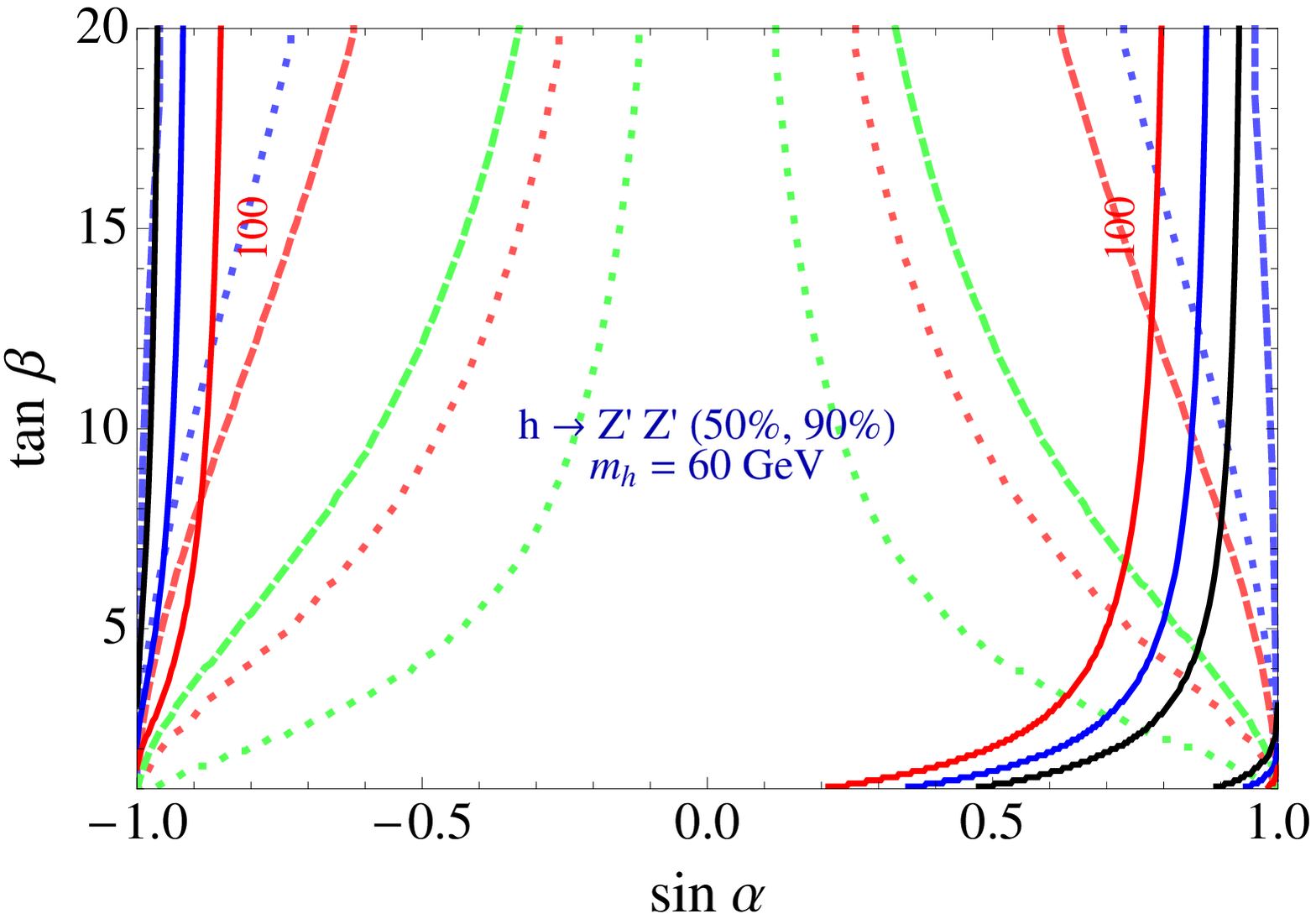} ~~~~~
\includegraphics[height=0.325\textwidth]{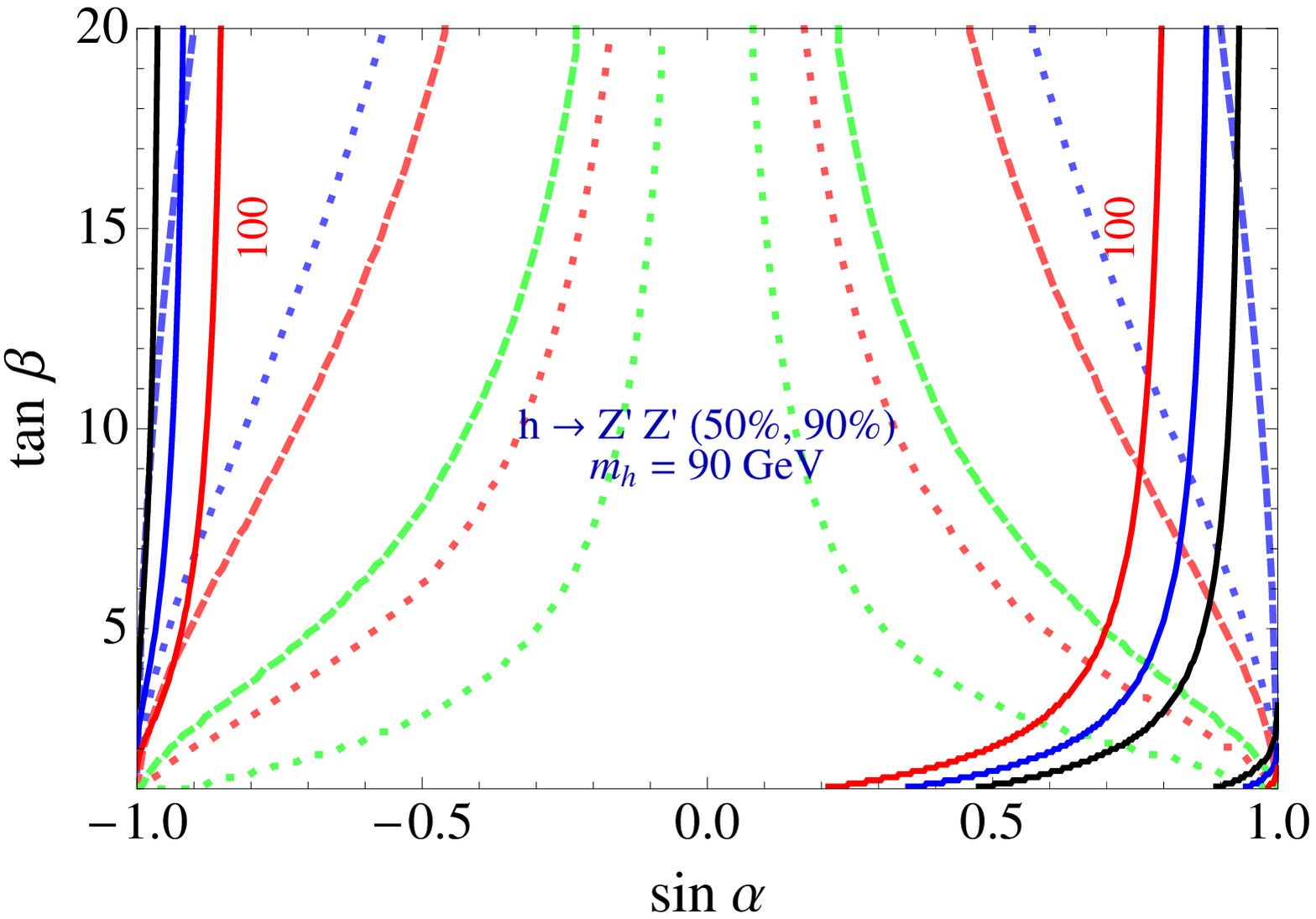}\\
~~~~~~ (a) ~~~~~~~~~~~~~~~~~~~~~~~~~~~~~~~~~~~~~~~~~~~~~~~~~~~~~~~~~~~~~~~~~~~~~~~~~~~~~~~ (b)
\end{center}
\caption{
Boundary where the light Higgs decay is dominated by $Z'Z'$ channel ($50\%$) for (a) $m_h = 60 ~\gev$ and (b) $m_h = 90 ~\gev$.
Dotted (Dashed) curves are where the branching ratio is $50 \%$ ($90 \%$).
Black solid curves are the LEP bound [as Fig.~\ref{fig:LEP} (b)] for the given $m_h = 60$, $90 ~\gev$.
Choices of $\delta \tan\beta$ and color code are the same as Fig.~\ref{fig:Darkdecays}.
}
\label{fig:LightToDark}
\end{figure*}

Figure~\ref{fig:LightToDark} shows the parameter region for $m_h = 60$ and $90 ~\gev$ in which $h \to Z'Z'$ dominates the light Higgs decay with $50 \%$, $90 \%$ of the total $h$ decay.
The $Z'Z'$ mode can dominate the light Higgs decay in a similar manner in which the Higgs decay to the weak vector bosons would dominate in the SM if the Higgs mass were sufficiently heavy.
This originates from the enhancement from the longitudinal polarization of vector bosons when they are boosted.

In accordance with expectation from Eq.~\eqref{eq:lightHiggsRatio}, Fig.~\ref{fig:LightToDark} shows that the $h \to Z'Z'$ dominates in a larger region of parameter-space with larger $m_h$ and larger $\delta \tan\beta$.
Especially, in the $\sin\alpha \sim \pm 1$ and sufficiently large $\tan\beta$ region, the $h$ decay is almost entirely into $Z'Z'$ as indicated by $90 \%$ dashed curves.

We note that for $\alpha \approx 0$ case, in which the lighter one would be the SM-like Higgs, the heavier one would dominantly decay into $Z' Z'$.
For some discussions of a Higgs decaying into $Z'Z'$ in different contexts, see Refs.~\cite{Gopalakrishna:2008dv,Lebedev:2011iq}.

The $Z'$ will then decay into fermions with a partial decay width, when fermion masses are neglected \cite{Davoudiasl:2012qa}.
\beq
\Gamma(Z' \to f \bar f) \simeq \frac{N_C}{48\pi} \eps_Z^2 g_Z^2 \left(g'^2_{V f} + g_{A f}^2\right) m_{Z'} ,
\eeq
where $g_{A f} \equiv -T_{3f}$ and $g'_{V f} \equiv T_{3f} - 2Q_f \left(\sin^2\theta_W - (\eps/\eps_Z)\cos\theta_W\sin\theta_W \right)$.  $N_C = 3\ (1)$ for quarks (leptons).
The branching ratio of $Z'$ into the charged leptons (which we call $x_e$ and $x_\mu$) depends on $\eps / \eps_Z$ as well as nontrivial hadronic decays.
It was shown in Ref. \cite{Thomas} that the individual lepton branching ratios for dark photons (i.e. a model with only kinetic mixing $\eps$ for the interaction) vary, roughly, between $10 \%$ and $40 \%$ over most of the mass range.
This branching ratio would change if there are other light hidden sector particles that $Z'$ can decay into.
Also, our dark $Z$ (i.e. a model with both kinetic mixing and $Z$-$Z'$ mass mixing parametrized by $\eps_Z$) has different branching ratios in general.
We assume no displaced vertex and take $x_\mu = 10 \%$ in the following sketchy analysis.

The light Higgs might be detectable by a similar resonance search for the heavier Higgs decaying into $Z'$ in Sec.~\ref{sec:125bounds} although a feasibility study might be necessary.
Very recently, ATLAS has looked for prompt ``lepton-jets'' at $7 ~\tev$ \cite{Aad:2012qua}.
A lepton-jet is a final state consisting of collimated muons or electrons.
In one of their analyses, with results consistent with the SM, they look for pairs of lepton-jets, each with two or more muons (the rest of the event is ignored).   Although they do consider a specific Hidden Valley model as an example, their results will apply to our case.
They only consider $Z'$ masses of $300$ and $500 ~\mev$, and the results are relatively insensitive to the choice of masses.   The resulting $95\%$ CL upper bound on the cross section times branching ratio is $17 ~\fb$ for $m_{Z'} = 300 ~\mev$ and $19 ~\fb$ for a $m_{Z'} = 500 ~\mev$. For our ballpark estimate, we will take the bound to be $20 ~\fb$ without specifying the $Z'$ mass.

The SM cross section for Higgs production through gluon fusion is about $37\ (30,\ 24) ~\pb$ for the Higgs mass of $80\ (90,\ 100) ~\gev$ at $7 ~\tev$ \cite{LHCHiggsProduction}.
In our case, the rate is multiplied by the factor $(\cos\alpha / \sin\beta)^2$ of Eq.~\eqref{eq:hcoupling}.
As shown in Fig.~\ref{fig:LightToDark}, the $h$ can decay into $Z'Z'$ dominantly in most of the parameter-space of interest.
If we ignore the $\tan\beta$ and $\sin\alpha$ dependence and take $\br (h \to Z'Z') \approx 1$, the light Higgs $h$ will appear as two muon-jets in the ATLAS analysis \cite{Aad:2012qua}.   Taking $m_h = 80 ~\gev$, the cross section times branching ratio for $pp \to h \to Z'Z' \to 2 ~\text{muon-jets}$ is bounded as
\beq
(37\ {\rm pb}) \left(\frac{\cos\alpha}{\sin\beta}\right)^2 x_\mu^2 ~ \lsim ~ 20\ {\rm fb} .
\eeq

For relatively large $\tan\beta$, this gives $\sin\alpha > 0.97$ for a choice of $x_\mu = 0.1$.   As $\tan\beta$ approaches $1$, this changes to $\sin\alpha > 0.99$.
As the $h$ mass varies, these numbers change slightly, but it is clear that one is forced into a fairly small area of parameter-space.    Although these values of $\sin\alpha$ might seem fine-tuned, one should note that it varies from $4$ to $7$ in terms of $\tan\alpha$ as $\sin\alpha$ varies from $0.97$ to $0.99$.
   
Given the level of our rather sketchy estimate, we do not take these bounds too seriously, but it is indicative that a sophisticated analysis can potentially shrink the allowed region considerably (unless the effect was discovered).

\section{\boldmath Decays and detection of Charged Higgs, $H^\pm$}
\label{sec:chargedHiggs}
\begin{figure}[bt]
\begin{center}
\includegraphics[height=0.325\textwidth]{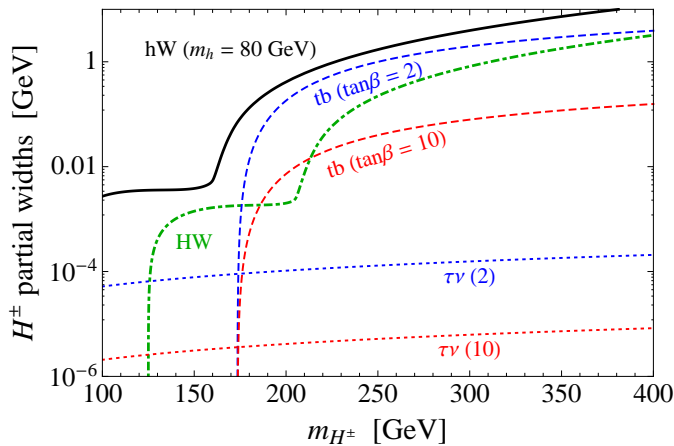}
\end{center}
\caption{
The partial decay widths for charged Higgs as a function of the charged Higgs mass.
The dashed lines are for the decay $H^+\to t\bar{b}$, with the upper (lower) line corresponding to $\tan\beta=2\ (10)$.   The dotted lines are similar for the decay $H^+\to\tau^+\nu$.  The solid line is for $H^+ \to hW^+$, which is either virtual or real, assuming $m_h = 80 ~\gev$ and $\cos^2 (\beta - \alpha) = 0.75$.
The dot-dashed line is for $H^+ \to HW^+$ with the $125 ~\gev$ SM-like Higgs.
The dominance of $hW$ decay mode is persistent for all $\cos^2 (\beta - \alpha)$ values allowed by the LEP over the mass range of interest:  $m_h \simeq 60 - 90 ~\gev$.
}
\label{fig:chargedHiggs}
\end{figure}

We have noted that the Dark 2HDM is unusual in that there is no pseudoscalar Higgs boson (it becomes the longitudinal component of the $Z'$) associated with the doublets\footnote{There is a pseudoscalar associated with a Higgs singlet, which remains decoupled as we assume no mixing between the doublets and singlet. When sizable mixing is introduced, the remaining pseudoscalar would have diluted coupling to the SM particles and it could be in principle detectable.}.
However, it certainly does have a charged scalar, $H^\pm$, which is orthogonal to the longitudinal component of the $W^\pm$.  The phenomenology of the charged Higgs in the ordinary Type I model was recently discussed in Ref.~\cite{Branco:2011iw}.

In the Type I model as well as in this Dark 2HDM, the coupling of the charged Higgs to fermions is suppressed by a factor of $\tan\beta$.
(Thus $B$ decay data, which pushes the charged Higgs mass above $300 ~\gev$ in the Type II model, is not stringent in the Type I model and in fact gives no bound for $\tan\beta > 2$ \cite{Branco:2011iw,Jung:2010ik}.)
Direct bounds on the charged Higgs mass from  LEP $H^+H^-$ searches give $m_{H^\pm} > 79 ~\gev$ assuming the charged Higgs decays only into $H^+ \to c \bar s$ and $\tau^+ \nu$ \cite{Searches:2001ac}.

If a charged Higgs is light enough, one can look for it in top quark decays, but as ATLAS data shows, the lower bound on $\tan\beta$ in the Type I model is fairly weak, varying from $1$ to $4$ in the mass range of $90 ~\gev < m_{H^\pm} < 140 ~\gev$ when the dominant decay mode is $\tau^+ \nu$ \cite{Aad:2012tj}.
For charged Higgs masses above about $180 ~\gev$, its primary decay mode is generally given by $H^+ \to t\overline{b}$, which is very difficult to detect.
For some discussion of relatively light charged Higgs ($m_{H^\pm} \gsim 90 ~\gev$) in ordinary 2HDMs, see Ref.~\cite{Aoki:2011wd}.

In the Type I model, one can also look for the decays $H^\pm \to \varphi W^\pm$ where $\varphi$ is either $h$, $H$ or $A$.
In the Dark 2HDM considered in this paper, however, there is a new and dramatic signature.
We have seen that the primary decay mode of the $h$ in most of the relevant parameter-space of our interest is into $Z'Z'$.
In this case, one can have $H^\pm \to h W^\pm \to Z' Z' W^\pm$.
Much of the time, the $Z'$ will decay into lepton pairs.
This will certainly be the dominant decay if the $H^\pm \to h W^\pm$ decay is kinematically a two-body decay, and would still be dominant even if the decay is kinematically three-body, since the competing decay will be suppressed by two powers of the $\tau$ mass.

The widths for the two and three-body decays of $H^\pm \to h W^\pm$ can be found in Refs.~\cite{Djouadi:1995gv,Akeroyd:1998dt}.
The two-body decay width is
\beq
\Gamma (H^\pm \to h W^\pm) = \frac{\cos^2(\beta - \alpha)}{16 \pi v^2} \frac{1}{m_{H^\pm}^3} \lambda^{3/2}(m_{H^\pm}^2,m_h^2,m_W^2) \label{eq:HhW}
\eeq
with $\lambda(x,y,z) \equiv x^2 + y^2 + z^2 - 2xy - 2yz - 2zx$.
It decreases with $m_h$ and increases with $\cos^2(\beta - \alpha)$.
Similarly, the decay width into the heavy Higgs ($125 ~\gev$) is given by
\beq
\Gamma (H^\pm \to H W^\pm) = \frac{\sin^2(\beta - \alpha)}{16 \pi v^2} \frac{1}{m_{H^\pm}^3} \lambda^{3/2}(m_{H^\pm}^2,m_H^2,m_W^2) ,
\eeq
and the decay width into $t \bar b$ is given by
\beq
\Gamma (H^\pm \to t \bar b) \simeq \frac{N_C m_{H^\pm}}{8 \pi v^2} \frac{m_t^2}{\tan^2\beta} \left(1 - \frac{m_t^2}{m_{H^\pm}^2}\right)^2
\eeq
when $m_b$ is ignored and $V_{tb} = 1$ is taken, and a similar expression is given for $\tau^+ \nu$ mode.

Figure~\ref{fig:chargedHiggs} shows the partial decay widths based on the above formulas for $H^+ \to t\bar{b}$, $H^+ \to \tau^+\nu$, $H^+ \to h W^+$, $H^+ \to H W^+$ assuming that $m_h = 80 ~\gev$ and $\cos^2(\beta - \alpha) = 0.75$.
(The $c \bar s$ has a smaller contribution than $\tau^+\nu$ as in the ordinary Type I model because of the small mass at the Higgs mass scale as the RG running is faster with color.)
For $m_h < m_{H^\pm} < m_W + m_h$ region, the three-body decay $H^+ \to h W^{+*} \to h f\bar{f'}$ (and similarly for $H$) is shown in the plot.

We see that the decay into $h W$ completely dominates for the entire range of $H^\pm$ masses in the plot ($m_{H^\pm} > 100 ~\gev$).
We checked that, for all our interested range of $m_h \simeq 60 - 90 ~\gev$ and $\cos^2(\beta - \alpha) \simeq 0.75 - 1$ that we discussed in Sec.~\ref{sec:constraints}, the $H^\pm \to h W^\pm$ decay mode keeps dominating over the other $H^\pm$ decay modes for the mass range of $m_{H^\pm} > 90 ~\gev$.
While the decay rate of the charged Higgs is the same as in ordinary Type I model, the difference comes from how the $h$ can decay, leading to more dramatic signatures.

The production cross section for charged Higgs bosons, for fairly large $\tan\beta$, is dominated by Drell-Yan pair production.
Each of these charged Higgs can mainly decay into $Z'Z' W^\pm$, giving four $Z'$s in the relevant parameter-space.
This may be looked for with lepton-jet searches \cite{Aad:2012qua} in a similar fashion we discussed in the previous section for $p p  \to h \to Z'Z'$ channel, with appropriate cuts and selections.    Note that the lepton-jet searches only require at least two muon-jets.
With four $Z'$s and a ${\cal O} (0.1)$ branching ratio for $Z' \to \mu^+\mu^-$, a substantial fraction of charged Higgs pairs will give a signal.    One can estimate that the current ATLAS bound of 20 fb is already covering a region of parameter-space, but it is clear that a much more detailed analysis is needed.  Given the unique signature of the model, with four $Z'$s and two $W$'s, a more targeted search could cover much more of the parameter-space.

In the Dark 2HDM, there is no tree-level vertex $H^\pm W^\mp Z'$ for the same reason that there is no $H^\pm W^\mp Z$ vertex at tree level.
Namely, if one goes into a basis in which only one Higgs doublet gets a vev, then the charged Higgs is entirely in the other doublet.   As a result, it can have no vev-dependent vertices.    Since $SU(2)_L$ is broken, this result will break down at one-loop, and thus $H^\pm \to W^\pm Z$ and $H^\pm \to W^\pm Z'$ can occur at one-loop.    
The effect of the Goldstone boson equivalence theorem, with a boosted $Z'$, would enhance this loop-suppressed decay potentially at meaningful level.

We discussed, in this paper, only the case of $m_{H^\pm} \gsim 100 ~\gev$.
It will be interesting and important to study how low the charged Higgs mass can be in this model while satisfying all the experimental and theoretical constraints, in view of the fact that most of the lower bounds on the $m_{H^\pm}$ were obtained based on the typical $t \bar b$, $c \bar s$ and $\tau^+ \nu$ modes.
It is also noteworthy that other variants of the Dark 2HDM [for example, with some of the SM fermions carrying nonzero $U(1)'$ charges] would give the similar $hW$ dominance since Eq.~\eqref{eq:HhW} is valid for all types of 2HDMs.

\section{Summary and Conclusions}
\label{sec:conclusion}
Although experiments at the LHC have recently found the SM-like Higgs boson with a mass of  $125 ~\gev$,
it is still important to search for other Higgs bosons, which occur in most extensions of the Standard Model.   In this paper, we considered a 2HDM with a new $U(1)$ gauge symmetry under which the SM particles are not charged (the Dark 2HDM), and discussed the physics of the other Higgs bosons, namely, the non-SM-like neutral Higgs and the charged Higgs scalar.
The additional Higgs doublet of this Dark 2HDM is charged by the $U(1)$ gauge symmetry, and this $U(1)$ plays the role of the $Z_2$ parity of the ordinary 2HDMs, thus forbidding tree-level FCNC.

Kinetic mixing between the $U(1)$ gauge groups will generally occur, as will mixing in the $Z$-$Z'$ mass matrix.    The possibility of a very light $Z'$ gauge boson has attracted increasing interest of late.    Such a light $Z'$ has been a subject of active experimental searches including the fixed target experiments at JLab in Virginia and at Mainz in Germany.
There are also searches using the decays from mesons at KLOE, BaBar, and Belle experiments.

The physics connecting the Higgs to a heavy $Z'$ (for example, the $Z'$ decays into the Higgs boson \cite{Agashe:2007ki,Barger:2009xg}) has been extensively studied, but connection of the Higgs to a very light $Z'$ has not been.    The light $Z'$ allows a very interesting scenario as the various Higgs boson decays can involve a light gauge boson which can decay into leptons with ${\cal O}(0.1)$ branching ratio along with possible enhancement from the longitudinal polarization.   Interestingly, in this model, the most important predictions/constraints on Higgs properties come from the properties of the $Z'$.  The $Z'$ is assumed to have a mass of ${\cal O} (1) ~\gev$ or less.
Such a light mass is the region most high energy collider analyses discard to avoid large SM backgrounds.

We considered the case in which the $125 ~\gev$ SM-like Higgs is the heavier scalar, and studied the light Higgs and charged Higgs in the Dark 2HDM.
After describing the complete model, we studied constraints arising from LEP bounds for invisible Higgs decays and width of the $Z$ as well as the $125 ~ \gev$ SM-like Higgs state seen at the LHC. We found that a model with just two Higgs doublets is excluded by the LEP invisible Higgs decay bounds combined with the precision electroweak physics, rare $B$ decay, etc. This issue can be resolved by introducing an additional Higgs singlet.

We then considered the production and detection of the light Higgs boson ($h$) whose dominant decay is into $Z'Z'$.   This leads to a remarkable and unusual signature.   The $Z'$ will decay with an appreciable branching fraction into a lepton pair, which (since the $Z'$ is so light) will form a collimated lepton jet.
A recent ATLAS experiment looked for events with two or more such ``muon-jets'', and their results show that significant constraints on the model would be possible if a sophisticated analysis with larger statistics were to follow.

The ATLAS muon-jets experiment also constrains the charged Higgs boson, which decays predominantly into $h W$  (either as a two-body or a three-body decay).
The charged Higgs in the Dark 2HDM can be very light compared to those in the ordinary 2HDMs.

In this paper, we limited ourselves to the case in which the $125 ~\gev$ SM-like Higgs is the heavier neutral Higgs, no mixing is present between the Higgs doublets and a singlet, and we focused on the region in which the light Higgs mass is in the range of $m_H / 2 \lsim m_h \lsim m_Z$.
Relaxing these limits will allow more studies.

\vspace{0.3cm}
\noindent
Acknowledgments:
This work was supported in part by the U.S. DOE under Grant No.~DE-AC05-06OR23177 (JLab) and in part by the NSF under Grant No.~PHY-1068008 (W\&M).
We thank P. Ferreira and R. Santos for useful discussions about the code that was used, after modification, to generate some numerical results, and ATLAS experimenters A. Hass, E. Strauss, and G. Watts for discussions about their muon-jet analysis.
HL thanks H. Davoudiasl, I. Lewis, and W. Marciano for many helpful discussions during the dark $Z$ projects.

\appendix
\section{\boldmath Constraints on $Z'$ properties}
\label{sec:Zprime}
In the $m_{Z'^0}^2 \ll m_{Z^0}^2$ limit, as shown in Sec.~\ref{sec:model}, the $Z$-$Z'$ mass-squared matrix is given by
\beq
M_{ZZ'}^2 = \mat{m_{Z^0}^2 & -\Delta^2 \\ -\Delta^2 & m_{Z'^0}^2} \simeq m_{Z^0}^2 \mat{1 & -\xi \\ -\xi & m_{Z'^0}^2 / m_{Z^0}^2} .
\eeq
We take the parametrization $\xi = \eps_Z + \eps \tan\theta_W$ which helps separate the $Z'$ interactions with $J_{NC}$ and $J_{em}$.
The $\eps$ comes from the kinetic mixing term, and the $\eps_Z$ originates from the $Z$-$Z'$ mass mixing from the Higgs ($\Phi_1$) that is charged under both $SU(2)_L \times U(1)_Y$ and $U(1)'$.

The mass-squared matrix then can be written as
\beq
M_{ZZ'}^2 \simeq m_Z^2 \mat{1 & -\left(\eps_Z + \eps \tan\theta_W\right) \\ -\left(\eps_Z + \eps \tan\theta_W\right) & m_{Z'}^2 / m_Z^2}
\eeq
taking $m_{Z'}^2 \simeq m_{Z'^0}^2$ which is realized for $\xi^2 \ll m_{Z'}^2 / m_Z^2$.
The $Z$-$Z'$ mass mixing parameter $\eps_Z$ is further parametrized by
\beq
\eps_Z \equiv \frac{m_{Z'}}{m_Z} \delta
\eeq
with, from Eqs.~\eqref{eq:mZ} - \eqref{eq:xi},
\bea
\delta &\simeq& \frac{\cos\beta \cos\beta_d}{\sqrt{1 - \cos^2\beta \cos^2\beta_d}}.
\eea
This would have been $\delta \simeq 1 / \tan\beta$ in the doublets only case ($\cos\beta_d = 1$).

The approximation $\delta \approx \cos\beta \cos\beta_d \sim 1 / (\tan\beta \tan\beta_d)$ would be valid in the limit where both $\tan\beta$ and $\tan\beta_d$ are large, and partly because of this reason, we use $\delta \tan\beta$ as our input instead of $\delta$ in the numerical analysis of this paper, which helps in estimating the Higgs singlet contribution.

Ignoring the higher order terms for small $\eps$ and $\eps_Z$ parameters (and their combination), we get
\bea
{\cal L}_\text{int} &=& - e J^\mu_{em} \hat A_\mu - g_Z J^\mu_{NC} \hat Z^0_\mu \\
&\simeq& - e J^\mu_{em} (A_\mu + \eps Z'^0_\mu) - g_Z J^\mu_{NC} (Z^0_\mu - \eps \tan\theta_W Z'^0_\mu) \label{eq:kinetic} \\
&\simeq& - e J^\mu_{em} (A_\mu + \eps Z'_\mu) - g_Z J^\mu_{NC} (Z_\mu + \eps_Z Z'_\mu) \label{eq:mass}
\eea
where Eq.~\eqref{eq:kinetic} is obtained after field redefinition to remove kinetic mixing term [Eq.~\eqref{eq:kineticMixing}] at leading order and Eq.~\eqref{eq:mass} is after the $Z$-$Z'$ mass-squared matrix diagonalization [Eq.~\eqref{eq:massMixing}].
We can see that, because of cancellation, there is no net $Z'$ coupling to the weak neutral current induced by the kinetic mixing $(\eps)$.

For a very light $Z'$, the kinetic mixing $\eps$ is constrained by various experiments including the electron beam dump, electron anomalous magnetic moment, and narrow resonance searches.
In particular, there are active searches using fixed target experiments at JLab (in Virginia) and at Mainz (in Germany) as well as the searches using the heavy meson decays at KLOE, BaBar, Belle experiments.
Very roughly, for the $10 ~\mev \lsim m_{Z'} \lsim 1 ~\gev$ range, the experimental bounds of $\eps^2 \lsim 10^{-5}$ are present.
With some combination of $\eps$ and $m_{Z'}$ (roughly, $\eps^2 \sim 10^{-6} - 10^{-5}$ and $m_{Z'} \sim 20 - 50 ~\mev$), the $Z'$ can explain the $3.6 \sigma$ deviation of the muon $g-2$.
(See Ref.~\cite{Davoudiasl:2012qa} for details.)
The $\eps$ basically parametrizes the vector coupling of the $Z'$ as it dominates the coupling to the electromagnetic current.
A good summary of the constraints and sensitivities from various experiments in the $\eps^2 - m_{Z'}$ parameter-space can be found in Ref.~\cite{McKeown:2011yj}.

The $Z'$ interaction to the weak neutral current is enabled by the $Z$-$Z'$ mass mixing $\eps_Z$, and it controls the axial coupling of the $Z'$ as it dominates the coupling to the weak neutral current.
The axial couplings expand the phenomenology of the light $Z'$ into more areas including the low energy parity violation and the enhancement of the production for the boosted $Z'$.
They were studied in Refs.~\cite{Davoudiasl:2012ag,Davoudiasl:2012qa,BNL}.
They include low-energy parity violation, rare $K$ decays, and rare $B$ decays.
Also, bounds from the Higgs decay exists as the presence of the $H \to Z Z'$ and $H \to Z' Z'$ modes should be still consistent with the LHC data which is consistent with the $125 ~\gev$ SM Higgs property.
Typically, $\delta \lsim 10^{-2} - 10^{-3}$, depending on various conditions, is expected to satisfy all the constraints.
There are some caveats and dependencies on other parameters about this bounds, and we will take $\delta \lsim 10^{-2}$ as a firm upper bound that we should satisfy in this paper.
This is quite a small quantity that cannot be achieved with the pure doublets case as $1 / \tan\beta$ cannot be too small (as discussed in Sec.~\ref{sec:3B}), which is an important reason that this model needs a Higgs singlet.


\end{document}